\documentclass[journal]{IEEEtran}
\usepackage{graphicx, nicefrac,subfigure,amsfonts, amsmath,amssymb,color,array,booktabs}
\usepackage{algorithm,algpseudocode,cite,epstopdf,enumitem}
\usepackage{stmaryrd,url}
\allowdisplaybreaks

\def\bbe{\boldsymbol{\beta}}

\def\bxi{\boldsymbol{\xi}}

\def\bPi{\boldsymbol{\Pi}}

\def\bom{\boldsymbol{\omega}}
\def\bGa{\mathbf{\Gamma}}

\def\bLa{\mathbf{\Lambda}}
\def\bXi{\mathbf{\Xi}}

\def\({\left(}
\def\){\right)}
\def\[{\left[\,}
\def\]{\,\right]}

\def\0{\boldsymbol{0}}
\def\1{\boldsymbol{1}}
\def\a{\mathbf{a}}
\def\A{\mathbf{A}}

\def\b{\mathbf{b}}
\def\B{\mathbf{B}}

\def\c{\mathbf{c}}
\def\C{\mathbf{C}}
\def\bC{\mathbb{C}}

\def\D{\mathbf{D}}

\def\H{\mathbf{H}}

\def\I{\mathbf{I}}

\def\J{\mathbf{J}}

\def\N{\mathbf{N}}

\def\P{\mathbf{P}}

\def\Q{\mathbf{Q}}

\def\s{\mathbf{s}}
\def\S{\mathbf{S}}

\def\T{\mathbf{T}}

\def\U{\mathbf{U}}

\def\v{\mathbf{v}}
\def\V{\mathbf{V}}

\def\y{\mathbf{y}}
\def\Y{\mathbf{Y}}

\def\Z{\mathbf{Z}}

\def\vec{{\mathrm{vec}}}

\def\diag{\mathrm{diag}}

\newtheorem{theorem}{Theorem}

\newtheorem{corollary}{Corollary}
\newtheorem{definition}{Definition}
\newtheorem{remark}{Remark}

\def\cO{\mathcal{O}}
\def\Hls{\H_\mathrm{LS}}

\def\tY{\pmb{\mathcal{Y}}}
\def\tX{\pmb{\mathcal{X}}}

\title{Algebraic Channel Estimation Algorithms for FDD Massive MIMO systems}
\author{Cheng Qian, \emph{Member, IEEE,} Xiao Fu, \emph{Member, IEEE}, and Nikolaos D Sidiropoulos, \emph{Fellow, IEEE} 
\thanks{
	Conference version of part of this work has been submitted to IEEE SPAWC 2019  \cite{QiFuSi:SPAWC2019}. This work was supported in part by the National Science Foundation under project NSF ECCS 1808159 and NSF ECCS 1608961. \par
	C. Qian and N. D. Sidiropoulos are with the Department of Electrical and Computer Engineering, University of Virginia, Charlottesville, VA 22904 USA (e-mail: alextoqc@gmail.com, nikos@virginia.edu).\par
	X. Fu is with the School of Electrical Engineering and Computer Science, Oregon State University, Corvallis, OR 97331 (xiao.fu@oregonstate.edu). 
}	
}

\begin{document}

\maketitle

\begin{abstract}
	We consider downlink (DL) channel estimation for frequency division duplex based massive MIMO systems under the multipath model. Our goal is to provide fast and accurate channel estimation from a small amount of DL training overhead.
	Prior art tackles this problem using compressive sensing or classic array processing techniques (e.g., ESPRIT and MUSIC). However, these methods have challenges in some scenarios, e.g., when the number of paths is greater than the number of receive antennas. Tensor factorization methods can also be used to handle such challenging cases, but it is hard to solve the associated optimization problems. In this work, we propose an efficient channel estimation framework to circumvent such difficulties.
	Specifically, a structural training sequence that imposes a tensor structure on the received signal is proposed. We show that with such a training sequence,  the parameters of DL MIMO channels can be provably identified even when the number of paths largely exceeds the number of receive antennas---under very small training overhead. Our approach is a judicious combination of Vandermonde tensor algebra and a carefully designed conjugate-invariant training sequence. Unlike existing tensor-based channel estimation methods that involve hard optimization problems,  the proposed approach consists of very lightweight algebraic operations, and thus real-time implementation is within reach.
	Simulation results are carried out to showcase the effectiveness of the proposed methods.
\end{abstract}


\section{Introduction}
\IEEEPARstart{M}{assive} MIMO promises significant performance gains in terms of spectral efficiency, reliability and security over the existing communication systems \cite{larsson2014massive,heath2016overview}. 
However,  realizing many of these advantages in practice hinges on accurate estimation of the channel state information (CSI), which affects the performance of transmit beamforming at the transmitters and decoding accuracy at the receiver.

Previously, much attention has been devoted to the time division duplex (TDD) protocol, where channel reciprocity can be invoked to estimate the downlink (DL) CSI from uplink (UL) training. However, this convenient property does not hold under the frequency division duplex (FDD) protocol, where UL and DL channels are in different frequency bands, with generally different propagation characteristics. Hence, the DL channel is different from the UL one, and it must be estimated by the receiver and then fed back to the transmitter. On the other hand, FDD offers uninterrupted full-duplex transmission, and relaxed amplification and synchronization requirements which are critical factors affecting service and deployment costs. Hence it is of great interest to come up with lightweight training and feedback strategies that require few resources.

To alleviate the heavy burden of DL training and UL feedback, one possible way is to reduce the effective channel parameters by considering a specular multipath channel comprising a few dominant paths, each characterized by direction-of-arrival (DOA), direction-of-departure (DOD) and channel gain \cite{heath2016overview,alkhateeb2014channel,qc2018,3gpp_stand}.
Such a channel model is effective under certain conditions, e.g., when the base station (BS) antenna array is mounted on top of a tall building or cellular tower, such that the number of local scatterers is limited. 
In addition, when the carrier frequency is lifted to the millimeter wave regime, due to the severe path loss, only a few specular reflections reach the other end of the link \cite{heath2016overview,xie2016overview,jarvis}.
Thus, the channel tends to exhibit a sparse structure in the angular domain. This allows for channel modeling using only DOA, DOD and channel gain. Under this model, the channel estimation problem for uniform transmit/receive arrays is related to multidimensional harmonic retrieval problems in classical array processing which have been well-studied in the past few decades \cite{van2004optimum}.
Array processing algorithms (such as maximum likelihood \cite{stoica1989music} and subspace based approaches \cite{rao1989performance,roy1989esprit}) can be employed to estimate multipath parameters. 
These methods are good fit for TDD systems but not for FDD. 
The reason is that array processing methods require a large array aperture for parameter estimation, where the array size should be greater than the number of paths in general---which is relatively easy to be satisfied in UL channels since the BS typically has many more antennas than the mobile station (MS), especially in massive MIMO scenarios. However, in FDD systems, the DL and UL channels have to be estimated separately. When the number of DL paths is larger than the number of receive antennas at the MS, conventional array processing methods will not work \cite{wax1989unique,qc2018}.


The above problem might be tackled by using compressive sensing (CS) methods. With a limited number of paths, the channel exhibits a sparse pattern in the angle domain, and thus channel estimation can be recast as a sparse regression problem \cite{jarvis,berger2010application,jomp,lee2016channel,he2016pilot, han2019fdd,han2019efficient,alevizos2018limited}. Many CS based algorithms have been developed. The authors of \cite{jarvis} employed CS for channel estimation  and proved that if both BS and MS are equipped with uniform linear arrays (ULAs), a MIMO channel with dimension $M_r\times M_t$ can be recovered from $\cO(K\log (M_rM_t/K))$ training samples with high probability, where $K$ is the number of paths, and $M_r$ and $M_t$ are the number of antennas at MS and BS, respectively. Since then a series of CS based techniques were proposed to enhance the channel estimation performance such as \cite{lee2016channel,jomp}.
The CS-based approach is elegant and works to some extent, but it also faces some challenges. CS-based methods rely on a discretized angle dictionary for parameter estimation, which is usually a very `fat' matrix with many coherent columns. This may lead to unsatisfactory performance for sparse recovery. Since DOAs/DODs are continuous variables in space, how to alleviate the performance loss caused by angle discretization is another crucial issue. One way is to use gradient descent \cite{yang2012off} or Newton's method \cite{qian2018robust,han2019efficient,han2019fdd} to refine the angles. But this involves additional optimization and increases the complexity.

There are also matrix completion (MC) techniques employed for multipath channel estimation, e.g., \cite{shen2015joint,li2018millimeter}. 
Fang \emph{et al.,} \cite{li2018millimeter} employed MC to solve the channel estimation problem of a millimeter wave system with a single radio frequency chain, where they assumed that the number of dominant paths is much smaller than the number of the transmit and receive antennas. This method requires multiple communications between the BS and MS to collect enough data over time to form a low-rank data matrix. Such a protocol implicitly assumes that MS and scatterers remain static. 
Furthermore, the overall training overhead is still high and solving the MC problem is a non-trivial task in terms of computational complexity. 
Note that in the current FDD systems, the MS never communicates with the BS for DL channel estimation; instead, the BS acts more like a radio station and only broadcasts training sequences and its basic service information.

Another way to reduce the computational burden on the mobile end is to exploit the so-called \textit{spatial reciprocity} \cite{han2019fdd,han2019efficient,hugl2002spatial,xie2017unified}. In this line of work, it is assumed that the UL and DL channels share the same propagation paths, and thus UL channel estimation yields important information for the DL channel as well. In this way, the DL estimation burden is shifted to the base station, which is anyway responsible for estimating the UL channel(s). This approach requires that the UL and DL operate on close-by carrier frequencies over similar bandwidths.
The challenge is that in many FDD systems, the DL channel can have a much wider bandwidth using multiple carriers in different bands, whereas the UL channel is usually on a single carrier/band. This causes a wide frequency separation (e.g., 1 GHz) between the DL and UL channels, which may then exhibit very different propagation characteristics\footnote{See 5G UL and DL frequency allocations in \url{https://www.everythingrf.com/community/5g-nr-new-radio-frequency-bands} and \url{https://en.wikipedia.org/wiki/5G_NR_frequency_bands}.}.

In \cite{qc2018}, an FDD massive MIMO system was considered with both BS and MS equipped with dual-polarized antennas. It has been shown that there is a hidden tensor structure in the received training data, and effective tensor factorization algorithms were proposed to estimate the multipath parameters.  
However, the techniques and parameter identifiability results therein are enabled by the special structure of dual-polarized multipath channels---how to generalize the technique to handle general multipath channels is unclear. In addition, the method in \cite{qc2018} is realized using computationally heavy optimization algorithms, which may not be realistic for mobile phones whose computational power is rather limited. 

In this paper, we consider parameter estimation for general specular multipath channels. We aim to provide effective estimation schemes that entail very low DL training overhead and low complexity.
Our detailed contributions are summarized as follows:
\begin{itemize}[leftmargin=3mm]
	\item \textbf{Short Training Sequence Design} 
	\item[] We propose a new training sequence with a conjugate symmetric structure for FDD massive MIMO systems. As we will see, our judicious design enables simple and effective channel estimation with very low training overhead.

	
	\item \textbf{Low complexity Algorithm}
	\item[] We show that by using the proposed training sequence, the received data can be transformed to a low-rank tensor, and thus channel estimation can be recast as low-rank tensor decomposition. Two simple algebraic methods are then devised for channel estimation.

	\item \textbf{Identifiability Analysis}
	\item[] We analyze the multipath parameter identification problem for massive MIMO. We show that under mild conditions, all parameters of the channel are identifiable using the proposed training sequence and algorithms. 
	
\end{itemize}

A short conference version of this work has been submitted to the IEEE SPAWC 2019 workshop \cite{QiFuSi:SPAWC2019}. This journal version includes a more advanced and accurate estimation method, fleshed out analysis, and more comprehensive experiments.

The remainder of this paper is organized as follows. 
In Section II we describe the signal model and multipath channel estimation problem. 
The major contributions of the paper appear in Sections III and IV; the former explains the design of a novel training sequence for frugal DL training, and the latter presents a computationally efficient channel estimation algorithm which incorporates the designed training sequence. Identifiability results are also provided in Section IV. Section V presents an improved channel estimator which incorporates the method proposed in Section IV and a root-finding technique to achieve higher estimation accuracy. Section VI presents our simulation results, and Section VII summarizes our conclusions.

\noindent\textbf{Notation:} Throughout the paper, superscripts $(\cdot)^T$, $(\cdot)^*$, $(\cdot)^H$, $(\cdot)^{-1}$ and $(\cdot)^\dagger$ represent transpose, complex conjugate, Hermitian transpose, matrix inverse and pseudo inverse, respectively. We use $|\cdot|$, $\|\cdot\|_F$, $\|\cdot\|_2$ and $\|\cdot\|_1$ for absolute value, Frobenius norm, $\ell_2$-norm and $\ell_1$-norm, respectively; $\hat a$ denotes an estimate of $a$, $\text{diag}(\cdot)$ is a diagonal matrix holding the argument in its diagonal, $\text{vec}(\cdot)$ is the vectorization operator and $\angle(\cdot)$ takes the phase of its argument; $[\cdot]_i$ is the $i$th element of a vector, $[\S]_{i,j}$ is the $(i,j)$ entry of $\S$, and $\s_{r,k}$ is the $k$th column of $\S_r$. Symbols $\otimes,\odot, \circledast \text{ and } \circ$ denote the Kronecker, Khatri-Rao, element-wise, and outer products, respectively; $[\S]_{i:j,m:n}$ extracts the elements in rows $i$ to $j$ and columns $m$ to $n$, $[\S]_{:,i:j}$ extracts the elements in the columns $i$ to $j$ and $[\S]_{i:j,:}$ extracts the elements in the rows $i$ to $j$. $\I_m$ is the $m\times m$ identity matrix and $\0_{m\times n}$ is the $m\times n$ zero matrix.

\section{Signal Model and Problem Statement}

\subsection{Channel Model}
We consider the DL of a FDD massive MIMO system, where a BS with $M_t$ transmit antennas sends signals to the MS that is equipped with $M_r$ receive antennas. After collecting $N$ temporal samples, the received data matrix at the MS is
\begin{align}\label{y}
\Y &= \H\S + \N
\end{align}
where $\H\in\bC^{M_r\times M_t}$ is the DL channel matrix, $\S\in\bC^{M_t\times N}$ is the training signal matrix, $\N\in\bC^{M_r\times N}$ is i.i.d. circularly symmetric complex Gaussian noise with mean zero and covariance $\sigma^2\I_{M_r}$. 
When the BS employs a transmit array with many antennas and the carrier frequency goes to 60 GHz, it is reasonable to assume that there are a few scatterers between the transmitter and receiver \cite{heath2016overview,3gpp_stand}. Under this assumption, the channel $\H$ is modeled as 
\begin{align}\label{H}
\H = \A_r\diag\big(\bbe\big)\A_t^H
\in\bC^{M_r\times M_t}
\end{align}
where
\begin{align*}
\A_r &= \begin{bmatrix} \a_r(\theta_{r,1},\phi_{r,1}) & \cdots & \a_r(\theta_{r,K},\phi_{r,K}) \end{bmatrix} \\
\A_t &= \begin{bmatrix} \a_t(\theta_{t,1},\phi_{t,1}) & \cdots & \a_t(\theta_{t,K},\phi_{t,K}) \end{bmatrix} \\
\bbe &=
\big[
\beta_1~ \cdots~\beta_K
\big]^T.
\end{align*}
In the above,  $K$ is the number of paths, $\beta_k$, $\a_t(\theta_{t,k},\phi_{t,k})$ and $\a_r(\theta_{r,k},\phi_{r,k})$ denote the gain, transmit and receive steering vectors of the $k$th path, respectively, where $\{\theta_{r,k},\phi_{r,k}\}$ are the azimuth and elevation angles of DOA and $\{\theta_{t,K},\phi_{t,K}\}$ are the azimuth and elevation angles of DOD. We assume that the BS is equipped with an $M_x\times M_y$ element  uniform rectangular array (URA), and the MS has a small uniform linear array (ULA) with $M_r$ antennas. In this case, the total number of transmit antennas is $M_t=M_xM_y$. The $k$th steering vector of the MS is
\begin{align*}
 \a_{r,k} = 
\begin{bmatrix}
1 & e^{j\omega_{r,k}} & \cdots & e^{j(M_r-1)\omega_{r,k}}
\end{bmatrix}^T
\end{align*}
and the steering vector at the BS is
\begin{align*}
 \a_{t,k} = \a_{y,k} \otimes \a_{x,k}
\end{align*}
where  $\omega_{r,k}=2\pi d\sin(\theta_{r,k})/\nu$, $[\a_{x,k}]_{l_x} = e^{j\omega_{x,k}}, l_x = 0,\cdots,M_x-1$ and $[\a_{y,k}]_{l_y} = e^{j\omega_{y,k}},l_y=0,\cdots,M_y-1$
with $\omega_{x,k}=2\pi (l_x-1)d\sin(\phi_{t,K})\cos(\theta_{t,K})/\nu$ and $\omega_{y,k} = 2\pi (l_y-1)d\sin(\phi_{t,K})\sin(\theta_{t,K})/\nu$. Here, $\nu$ is the wavelength, and $d$ is the inter-element spacing distance between two adjacent antennas, which is assumed to be smaller than or equal to half-wavelength.

\subsection{Problem Statement and challenges}
In an FDD system, the DL and UL channels are operated in different frequency bands, so the MS must estimate the DL channel first and then feed it back to the BS through a low-rate UL channel, where the number of feedback bits is limited. If the dimension of the channel is large, it is impractical to feed back the whole channel matrix. 
A more practical and economical way is to estimate and feed back the key parameters such as DOAs, DODs and path-losses that characterize the DL channel.

In practice, if the training sequences are orthogonal and both receive and transmit antennas are ULAs/URAs, the problem of estimating multipath parameters actually belongs to a class of multidimensional harmonic retrieval problems, which has been well-studied during the past few decades \cite{liu2002almost,liu2,nion2010tensor,qian2018robust}. To be specific, when $\S\S^\H=\I_{M_t}$, one can first estimate $\H$ via
$$  \hat{\H} = \Y\S^H = \A_r\diag(\bbe)(\A_y\odot\A_x)^H + \N\S^H$$
which is a 3-D harmonic retrieval (HR) model \cite{liu2002almost,liu2,nion2010tensor,unitaryesprit,smoothESPRIT,sorensen2017multidimensional}.
Then, the key parameters can be estimated from $\hat{\H}$ via various approaches such as \cite{liu2002almost,liu2,nion2010tensor,unitaryesprit,smoothESPRIT,sorensen2017multidimensional,hua1990matrix}, even when $K\gg M_r$. 
However, the 3d-HR approach is computationally expensive, and using an orthogonal $\S$ means that $N\geq M_t$ has to be satisfied. When the number of transmit antennas is large, this inevitably leads to high training overhead---which is undesired in massive MIMO systems, especially under mobility, where agile channel estimation is need. 
When $N<M_t$ and the training signal $\S$ is non-orthogonal, i.e.,
$$\S\S^H\neq\I_{M_t},$$ 
the matched filtering output
\begin{align}\label{Hls}
\Hls = \H\S\S^H \neq \H
\end{align}
is no longer a good approximation of the original channel, even without any noise. Under such circumstances, estimating the channel parameters becomes very challenging, and only a few cases are known to be resolvable. 
One major challenge is identifiability. Based on the existing identifiability results for array processing \cite{wax1989unique,nehorai1991direction}, given $\Y$ and an unstructured $\S$, the number of paths that we can handle is about $(M_r-1)$. In other words, once $K\geq M_r$ which is the case in practical scenarios, the channel parameters may not be identifiable.
Even if $K<M_r$, conventional array processing methods can only identify $\A_t^H\S$ instead of $\A_t$, but how to efficiently estimate the DODs from this term is unclear. In \cite{qc2018}, an iterative optimization algorithm was proposed to estimate the DODs from a similar term, but the complexity of the algorithm may be too high for a practical commercial smart phone. 
When $K\geq M_r$, one may adopt CS based methods to estimate multipath parameters \cite{jarvis,jomp,lee2016channel}. However, sparse methods also face serious challenges.
Specifically, discretizing the angular space leads to sub-optimality and solving a large-scale sparse optimization with a semi-coherent dictionary is a challenge for practical implementation.

\begin{figure*}
	\centering
	\includegraphics[width=0.95\linewidth]{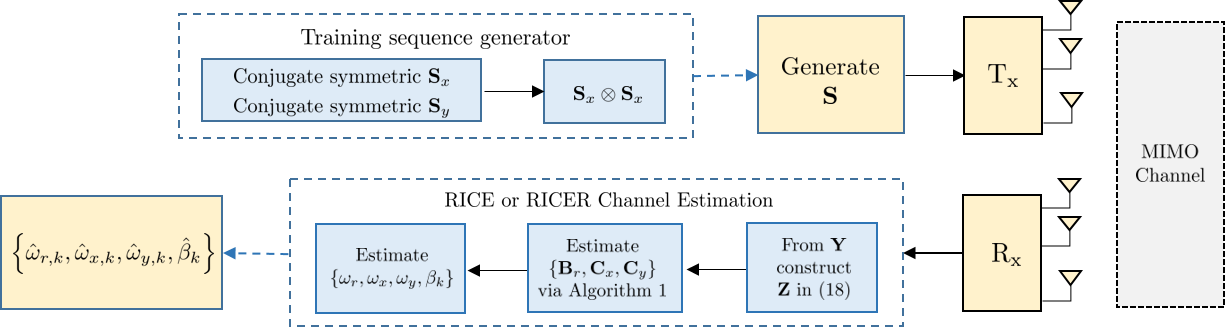}
	\caption{Proposed frame structure for downlink channel estimation.}
	\label{fig:configuration}
\end{figure*}

\section{Training Sequence Design}
This work consists of two components for channel estimation: training sequence design and channel parameter estimation. Fig. \ref{fig:configuration} shows the block diagram of the proposed system.
In this section, we will discuss the first part--training sequence design, which is critical for the subsequent channel estimation. We propose to design a ``tall'' training matrix $\S$ which has certain structure to overcome the difficulties mentioned above.

\subsection{Tensor Preliminaries}
To make the paper self-contained, we briefly present the definition of tensor rank and some useful theorems on the uniqueness of tensor decomposition in the following.

\begin{definition}\label{definition:tensor}
	\textit{(Canonical Polyadic Decomposition (CPD))}. A tensor is a multidimensional array indexed by three or more indices. Specifically, an third order tensor $\tX\in\bC^{I\times J\times K}$ that has three latent factor matrices $\{\A,\B,\C\}$ can be written as
	\begin{align*}
	\tX &= \sum_{f=1}^{F} [\A]_{:,f} \circ [\B]_{:,f} \circ [\C]_{:,f} \notag\\
	&:= \left\llbracket \A,\B,\C \right\rrbracket
	\end{align*}
	where $\A\in\bC^{I\times F}$ and $[\A]_{:,f} \circ [\B]_{:,f} \circ [\C]_{:,f}$ is a rank-1 tensor. The minimal such $F$ is the rank of tensor $\tX$ or the CPD rank of $\tX$ \cite{Sid2017}.
\end{definition}

\begin{definition}\label{definition:unfolding}
	(\textit{Unfolding}).	
	Tensor unfolding is obtained by taking the mode-$n$ slabs of the tensor (i.e., subtensors obtained by fixing the $n$th index of the original tensor), vectorizing the slabs, and then stacking all the vectors from left to the right into a matrix
\end{definition}

CPD factors $\tX$ into a sum of rank-one tensors.  
It is known that the CPD is unique under mild conditions, up to scaling and permutation of the $F$ components. This is referred to as ``essential uniqueness'' of the latent factors in the literature, and formally defined as follows. 
\begin{definition}
	(\textit{Uniqueness}).  Given a $N$-th order tensor $\tX = \llbracket\U_1,\cdots,\U_N\rrbracket$ of rank $F$, its CPD is essentially unique if the rank-one terms in the decomposition are unique, i.e., there is no other way to decompose $\tX$ for the given number of rank-1 terms.
	If
	$\tX = \llbracket\breve\U_1,\cdots,\breve\U_N\rrbracket$, for some $\{\breve{\U}_n\}_{n=1}^N$, then there exists a permutation matrix $\bPi$
	and diagonal matrices $\{\bXi_n\}_{n=1}^N$ such that
	\begin{align*}
	\breve\U_n = \U_n\bPi\bXi_n, \forall n=1,\cdots,N
	\end{align*}
	where $\prod_{n=1}^{N}\bXi_n = \I_F$.
\end{definition}

In some special cases, the factor matrices have special structure, e.g., they are Vandermonde. With this prior information, one can show stronger identifiability result. For example,
\begin{theorem} \cite{smoothESPRIT}\label{theorem:3wayVdmd}
	Consider a third-order tensor $\tX = \llbracket\A,\B,\C\rrbracket$, where $\A\in\bC^{I\times F}$, $\B\in\bC^{J\times F}$, $\C\in\bC^{K\times F}$, $\A$ is Vandermonde with distinct nonzero generators. Assume that $\B$ and $\C$ are drawn from an absolutely continuous distribution. If
	\begin{align*}
	F \leq \min\Big(  (I_{1}-1)J,~ I_2K  \Big)
	\end{align*}
	where $I_1\geq I_2$ and$I_2= I+1- I_{1}$ are chosen from
	\begin{align*}
	\{I_1,I_2\} = \arg\max_{\{I_1,I_2\}\in\mathbb{Z}^+}~ &\min\Big((I_{1}-1)J, I_2K\Big)
	\end{align*}
	then $\A$, $\B$ and $\C$ are essentially unique with probability one.
\end{theorem}

\subsection{Conjugate Flipped Structure}\label{section2b}
Our idea is to use a Vandermonde structure-enabled algebraic tensor factorization algorithm to recover an $M_r\times M_t$ channel matrix from the received signal matrix $\Y$ with dimension $M_r\times N$, where $N<M_t$. Then we use a simple algorithm to recover the DOAs and DODs. Both steps are enabled via a judiciously designed training sequence. Let us first show how to transform $\Y$ to a tensor. It follows from $(\A\otimes\B)^H(\C\odot\D) = (\A^H\C)\odot(\B^H\D)$ that by defining
\begin{align}\label{X}
\S = \S_y\otimes\S_x
\end{align}
$\Y$ can be written as
\begin{align}\label{Y}
\Y 
&= \A_r\diag\(\bbe\)\(\S_y^H\A_y\odot\S_x^H\A_x\)^H \notag\\
&\overset{\Delta}{=} \B_r\(\C_y\odot\C_x\)^H
\end{align}
where $\B_r=\A_r\diag\(\bbe\)$ with its $k$th column being $\b_{r,k} = \beta_k\a_{r,k}$, $\S_x\in\bC^{M_x\times N_x}$, $\S_y\in\bC^{M_y\times N_y}$, $\C_x = \S_x^H\A_x\in\bC^{N_x\times K}$ and $\C_y = \S_y^H\A_y\in\bC^{N_y\times K}$ with $N=N_xN_y$. Note that the scalar $\beta_k$ does not hurt the Vandermonde structure in $\a_{r,k}$, so $\B_r$ is Vandermonde.
According to Definitions 1 and 2, $\Y$ in \eqref{Y} is the matrix form of a third-order tensor with rank $K$ defined as
\begin{align}\label{tY}
\tY = \sum_{k=1}^K\b_{r,k}\circ\c_{x,k}^*\circ\c_{y,k}^* = \left\llbracket\B_r, \C_x, \C_y\right\rrbracket.
\end{align}

Before we continue, it is necessary to note that the essential uniqueness of tensor factorization makes the latent factors of a tensor identifiable under mild conditions \cite{Sid2017}. In our case, the latent factors are $\{\B_r, \C_x, \C_y\}$ and they are identifiable from $\tY$ up to column permutation and scaling ambiguity under some conditions. As we will see in Section IV, with the Vandermonde $\B_r$, these factor matrices can be efficiently identified by computing singular-value decomposition (SVD) of a small dimensional matrix, and hence avoiding the complicated optimization procedure as conventional tensor decomposition approaches do.
However, our target is not the factor matrices but the angles and path-losses contained therein. 
The estimation of DOAs is relatively simple because $\B_r$ is Vandermonde and we can estimate DOAs from the columns of $\B_r$. The difficulty here is the estimation of DODs and path-losses, where the former are contained in $\C_x$ and $\C_y$ while the latter are not even identifiable from standard tensor factorization approaches. 
In the following, we will show that by designing a specially structured training matrix $\S$, all the multipath parameters are identifiable from a simple algebraic method with identifiability guarantees. 


Assume that we have already identified $\C_x$ and $\C_y$. The remaining task is to identify azimuth and elevation angles from $\C_x$ and $\C_y$.
By definition, $\C_x=\S_x^H\A_x$ and $\C_y=\S_y^H\A_y$, so the designs of $\S_x$ and $\S_y$ are the same. To simplify the analysis, let us temporally remove the subscripts $x$ and $y$, and consider the design of $\S$ for an $M$-element ULA that has the steering vector as
\begin{align*}
	\a = 
	\begin{bmatrix}
		1& e^{j\omega} & \cdots & e^{j(M-1)\omega}
	\end{bmatrix}^T
\end{align*}
with $\omega\in\[-\pi,\pi\]$.
Assume that there is a training signal defined as
\begin{align}\label{barw1}
	\overline{\s}_l = 
	\begin{bmatrix}
		s_1 & s_2 & \cdots & s_{M+1-l} & \0_{l-1}
	\end{bmatrix}^T,~\forall l\geq 1
\end{align}
where $s_i$ is the symbol transmitted by the $i$th antenna element.
The inner product between $\a$ and $\overline{\s}_l$ is
\begin{align*}
	\overline{\s}_l^H\a = \sum_{m=1}^{M+1-l} s_m^*e^{j(m-1)\omega}
\end{align*}

Taking the conjugate of $\overline{\s}_l$ and then flipping its nonzero elements yields
\begin{align}\label{barw2}
	\underline{\s}_l = 
	\begin{bmatrix}
		s_{M+1-l}^* & \cdots & s_2^* & s_1^* & \0_{l-1}
	\end{bmatrix}^T
\end{align}
which leads to
\begin{align}\label{rie}
	\underline{\s}_l^H\a &= \sum_{m=1}^{M+1-l} s_{M+l-m}e^{j(m-l)\omega} \notag\\
		&= \sum_{m=1}^{M} (s_m^*e^{j(m-1)\omega})^*e^{j(M-l)\omega} \notag\\
		&= (\overline{\s}_l^H\a)^*a_{M+1-l}
\end{align}
where $a_{M+1-l}$ is the $(M+1-l)$th element of $\a$.
The above equation exhibits a ``conjugate'' rotational invariance (CRI) between $\overline{\s}_l^H\a$ and $\underline{\s}_l^H\a$, which is slightly different from the standard rotational invariance (as in, e.g., ESPRIT \cite{roy1989esprit}). The latter is usually built upon the forward and backward subarrays that are rotationally invariant by a factor of $e^{j\omega}$ while CRI does not rely on subarrays and its invariant factor is the $(M+1-l)$th element of the steering vector, i.e., $e^{j(M-l)\omega}$. Note that the insight of constructing CRI appeared in MIMO radar beamparttern design \cite{khabbazibasmenj2014efficient}, where a type of nonzero conjugately flipped waveform has been studied. 

Let us rewrite \eqref{rie} as
\begin{equation}\label{am}
	a_{M+1-l} = \frac{\overline{\s}_l^H\a}{(\underline{\s}_l^H\a)^*}.
\end{equation}
We see that it provides a way for estimating the phase contained in $a_{M+1-l}$, i.e., $(M-l)\omega$ which contains the target $\omega$. 
Nevertheless, for large $M$ and small $l$, $(M-l)\omega$ can be greater than $2\pi$, causing the so-called phase wrapping problem. Thus, we cannot find the exact $\omega$ from \eqref{am}. This is also the problem of \cite{khabbazibasmenj2014efficient}.
To solve the phase wrapping problem,  we need at least two adjacent elements of $\a$ since $\a$ is Vandermonde. In \eqref{am}, we have shown that a pair of $\overline{\s}_l$ and $\underline{\s}_l$ extracts the $(M+1-l)$th element of $\a$. 
Provided that there exists another pair denoted by $\{\overline{\s}_{l+1},\underline{\s}_{l+1}\}$ that takes the $(M-l)$ element of $\a$ out, we may obtain a Vandermonde vector which consists of $a_{M+1-l}$ and $a_{M-l}$, such that the phase $\omega$ can be estimated exactly through $\angle(a_{M+1-l}/a_{M-l})$. 
Based on this observation, we vary $l$ from 1 to $L$ and collect all $\{\overline{\s}_l\}$ and $\{\underline{\s}_l\}$ in, respectively,
\begin{subequations}
	\begin{align}
	\overline{\S} &= \[ \overline{\s}_{L} ~ \cdots ~ \overline{\s}_1 \] \label{over:w}\\
	\underline{\S} &= \[ \underline{\s}_{L} ~ \cdots ~ \underline{\s}_1 \]. \label{under:w}
	\end{align}
\end{subequations}
Then the following equality holds
\begin{align}\label{wa}
\underline{\S}^H\a = (\overline{\S}^H\a)^* \circledast \v  \in\bC^L
\end{align}
where $\v = \[\a\]_{M+1-L:M} = \[a_{M+1-L} ~ \cdots ~ a_M\]^T$ 
contains the last $L$ elements of $\a$, and hence it is Vandermonde. 

The estimation of $\omega$ is now much easier.
In practice, once $\underline{\S}^H\a$ and $\overline{\S}^H\a$ are identified\footnote{We will see later on how to identify them.}, we first estimate
\begin{align}\label{v}
\hat\v = \(\diag(\overline{\S}^H\a)^*\)^{-1} (\underline{\S}^H\a)
\end{align}
and then calculate the phase $\omega$ through 
\begin{align}\label{psi}
\hat\omega = \angle(\[\hat\v\]_{1:L-1}^H\[\hat\v\]_{2:L}).
\end{align}

\subsection{Design of $\S_x$ and $\S_y$}
Now let us return to the design of $\S_x$ and $\S_y$. 
We first talk about the design of $\S_x$.
To implement the above idea, the training signal must contain both $\overline{\S}$ and $\underline{\S}$.
One possible choice for $\S_x$ is
	\begin{align}\label{Si}
	\S_x &= \begin{bmatrix} \overline{\S}_x & \underline{\S}_x \end{bmatrix} \notag\\
	&=  \begin{bmatrix} \overline{\s}_{x,L} ~ \cdots ~ \overline{\s}_{x,1} & \underline{\s}_{x,L} ~ \cdots ~ \underline{\s}_{x,1} \end{bmatrix}  \in\bC^{M_x\times N_x}
	\end{align}
where $N_x=2L$, $\overline{\s}_{x,l}$ and $\underline{\s}_{x,l}$ are defined in \eqref{barw1} and \eqref{barw2}, respectively; $\overline{\S}_x$ and $\underline{\S}_x$ have the same definitions as \eqref{over:w} and \eqref{under:w}, respectively, and both of them have $L$ columns.

In Section \ref{section2b}, we show that the estimation of $\omega$ is only related to the spatial structure of $\overline{\S}$ and $\underline{\S}$ but not their values. This provides more freedom on choosing the values for $\S_x$ and $\S_y$. However, random $\overline{\S}$ and $\underline{\S}$ only guarantee the recovery of the phase but not the path loss. In order to estimate all the key parameters efficiently, we need one more constraint on choosing the values of $\S_x$ and $\S_y$.
We enforce the last elements in $\overline\s_{x,1}$ and $\overline\s_{y,1}$ to satisfy
\begin{align}\label{ss}
[\overline{\s}_{x,1}]_{M_x} [\overline{\s}_{y,1}]_{M_y} = 1.
\end{align}
Due to the conjugate symmetric property between $\overline{\S}$ and $\underline{\S}$, we also have $[\underline{\s}_{x,1}]_{1} [\underline{\s}_{y,1}]_{1} = 1$.

It is instructive to showcase the structure of $\S_x$ and $\S_y$ by examples. Let us consider a setting where $M_x=M_y=5$ and $N_x=N_y=4$. In this case, $\S_x$ and $\S_y$ are
\begin{align*}
\S_x =
\begin{bmatrix}
s_1 & s_1 & s_4^* & s_5^* \\ 
s_2 & s_2 & s_3^* & s_4^* \\ 
s_3 & s_3 & s_2^* & s_3^* \\ 
s_4 & s_4 & s_1^* & s_2^* \\
0   & s_5 & 0     & s_1^*
\end{bmatrix},\quad
\S_y =
\begin{bmatrix}
s_1 & s_1 & s_4^* & 1/s_5^* \\ 
s_2 & s_2 & s_3^* & s_4^* \\ 
s_3 & s_3 & s_2^* & s_3^* \\ 
s_4 & s_4 & s_1^* & s_2^* \\
0   & 1/s_5 & 0     & s_1^*
\end{bmatrix}
\end{align*}
where the first two columns in $\S_x$ are $\overline{\S}_x$ and the last two columns are $\underline{\S}_x$; similar to $\S_y$.
When $N_x\neq N_y$, for example $M_x=5, M_y=4$, we may choose
\begin{align*}
\S_x =
\begin{bmatrix}
s_1 & s_1 & s_4^* & s_5^* \\ 
s_2 & s_2 & s_3^* & s_4^* \\ 
s_3 & s_3 & s_2^* & s_3^* \\ 
s_4 & s_4 & s_1^* & s_2^* \\
0   & s_5 & 0     & s_1^*
\end{bmatrix},\quad
\S_y =
\begin{bmatrix}
s_2 & s_2 & s_4^* & 1/s_5^* \\ 
s_3 & s_3 & s_3^* & s_4^* \\ 
s_4 & s_4 & s_2^* & s_3^* \\
0   & 1/s_5 & 0   & s_2^*
\end{bmatrix}.
\end{align*}

\begin{remark}
	It is seen from the above examples that to construct $\S_x$ and $\S_y$, we only need to generate $\max(M_x,M_y)$ different symbols $\{s_i\}$.
	The minimum $L$ that guarantees the recovery of $\omega$ is 2, meaning that the minimum number of columns in $\S_x$ and $\S_y$ is $N_x=N_y=2L=4$. This also indicates that the minimum number of training samples for the above design is $N=N_xN_y=16$.
\end{remark}

\section{Channel Estimation}
In this section, we derive a computationally efficient channel estimator. We first explain the details on how to efficiently estimate the factor matrices $\{ \B_r,\C_x,\C_y \}$ in \eqref{tY}. Then we derive closed-form solutions for multipath parameter estimation. Finally, we claim uniqueness condition for the identification of these parameters.

\subsection{Identification of Factor Matrices}
According to Definition 2, the matrix unfolding of $\tY$ along its third dimension takes the form of
\begin{align}\label{Y3}
\Y_{(3)} =  \(\C_x^*\odot\B_r\)\C_y^H.
\end{align}
Since $\B_r$ is Vandermonde, the spatial smoothing technique is applicable to further expand the dimension of $\Y_{(3)}$. Specifically, defining a cyclic selection matrix $\J_{i_2}=\I_{M_x}\otimes[\0_{P_r\times i_2}~ \I_{P_r}~ \0_{P_r\times (M_r-i_2-P_r)}]$ and varying $i_2$ from 0 to $(Q_r-1)$, we have
\begin{align}\label{Z}
\Z &= 
\begin{bmatrix}
\J_{0}\Y_{(3)} & \cdots & \J_{Q_r-1}\Y_{(3)}
\end{bmatrix} \notag\\
&= \(\C_x^*\odot\B_1\) \( \B_2\odot\C_y^* \)^T \in \bC^{P_rM_x\times Q_rM_y}
\end{align}
where $P_r+Q_r = M_r + 1$.

Since $\B_1,\B_2$ are Vandermonde, given $\{P_r, Q_r\}$, we can follow \cite{smoothESPRIT,qc2018,fu2015factor} and employ an ESPRIT-like approach shown in Algorithm 1 to estimate $\B_r$, $\C_x$ and $\C_y$.
\begin{algorithm}
	\caption{Factor Matrices Estimation}
	\label{ESPRIT}
	\begin{algorithmic}[1]\vspace{.2em}
				
		\State Rearrange $\Z$ as $\tilde\Z = \(\B_1\odot\C_x^*\) \( \B_2\odot\C_y^* \)^T$ and calculate its SVD as $\tilde\Z = \U\bLa\V^H$.
		
		\State Choose the signal subspace as the $K$ principal singular vectors in $\U$, i.e., $\U_s = [\U]_{:,1:K}.$
		
		\State
		Define $\J_1 = [\I_{P_r-1},\0_{(P_r-1)\times 1}]\otimes\I_{M_x}$ and $\J_2 = [\0_{(P_r-1)\times 1},\I_{P_r-1}]\otimes\I_{M_x}$, and construct $\U_{s1}=\J_1\U_s$ and $\U_{s2}=\J_1\U_s$.
		
		\State 
		Let $\T$ be the left eigenvector matrix of $\U_{s1}^\dagger\U_{s2}$ and then calculate
		\begin{align*}
		 \B_1\odot\C_x^* &= \U_s\T \overset{\Delta}{=} \bGa_1 \\
		 \B_2\odot\C_y^* &= \(\T^{-1}[\bLa]_{1:K,1:K}[\V]_{:,1:K}^H\)^T \overset{\Delta}{=} \bGa_2.
		\end{align*}
		Now we can estimate $\C_x$ by choosing the first $M_x$ rows of $\bGa_1^*$ and $\C_y$ as the first $M_y$ rows of $\bGa_2^*$. 
		Finally, we estimate $\B_r$ as $\B_r = ((\C_y^*\odot\C_x^*)^\dagger\Y_{(1)})^T$,
		where
		\begin{align}\label{Y1}
		\Y_{(1)} = \(\C_y^*\odot\C_x^*\)\B_r^T.
		\end{align}
	\end{algorithmic}
\end{algorithm}

\subsection{DOA/DOD Estimation}
The factor matrices identified from Algorithm 1 suffer column permutation and scaling ambiguity, implying that the estimates of $\{\B_r,\C_x,\C_y\}$ are not exactly the original factors. Fortunately, this will not be an issue for angle estimation. Since the columns in $\{\B_r,\C_x,\C_y\}$ are paired with each other and the scaling ambiguity does not affect the array manifold structure, we can estimate the $k$th DOA and DOD from the $k$th columns of $\B_r$, $\C_x$ and $\C_y$, respectively.

Due to the one-by-one mapping between $\theta_{r,k}$ and $\omega_{r,k}$, estimating DOA is equivalent to estimate the phase $\omega_{r,k}$ that is calculated as
\begin{align}\label{omegar}
\hat\omega_{r,k} = \angle\( [\hat{\b}_{r,k}]_{1:M_r-1}^H [\hat{\b}_{r,k}]_{2:M_r}\),~k=1,\cdots,K
\end{align}
It is optional to estimate DOAs from
\begin{align}\label{doa}
\hat{\theta}_{r,k} = \arcsin\( \frac{\nu}{2\pi d}\hat\omega_{r,k} \).
\end{align}

The estimation of azimuth and elevation angles of DOD is different from DOA estimation due to the presence of $\S_x$ and $\S_y$.
To estimate them, it is necessary to know the phases contained in $\A_x$ and $\A_y$. 
Toward this end, let us consider the estimation of the phase $\omega_{x,k}$. Let $\hat{\overline{\C}}_{x}$ and $\hat{\underline{\C}}_{x}$ be submatrices of $\hat{\C}_x$ which contain the first and last half of the rows of $\hat{\C}_x$, respectively. 
According to the definition of $\S_x$ in \eqref{Si}, in the noiseless case, the explicit expressions for $\overline{\C}_{x}$ and $\underline{\C}_{x}$ are
\begin{subequations}
	\begin{align}
		\overline{\C}_{x} &= \overline{\S}_x^H\A_x \\
		\underline{\C}_{x} &=\underline{\S}_x^H\A_x.
	\end{align}
\end{subequations}
It follows from \eqref{v} that the top $N_x/2$ rows of the $k$th column in $\A_x$ equal to
\begin{align}
	\hat\v_{x,k} = \diag\(\hat{\overline{\c}}_{x,k}^*\)^{-1}\hat{\underline{\c}}_{x,k}.
\end{align}
The corresponding phase is then calculated via
\begin{align}\label{omegax}
\hat{\omega}_{x,k} = \angle\( [\hat{\v}_{x,k}]_{1:N_x/2-1}^H [\hat{\v}_{x,k}]_{2:N_x/2}\).
\end{align}
Similarly, by splitting $\hat\C_y$ into $\overline{\C}_{y}$ and $\underline{\C}_{y}$, we have 
\begin{align}\label{omegay}
\hat{\omega}_{y,k} = \angle\( [\hat{\v}_{y,k}]_{1:N_y/2-1}^H [\hat{\v}_{y,k}]_{2:N_y/2}\)
\end{align}
where $\hat\v_{y,k} = \diag\(\hat{\overline{\c}}_{y,k}^*\)^{-1}\hat{\underline{\c}}_{y,k}$.
It is optional to estimate the azimuth and elevation angles of DOD. If interested, we calculate them through 
\begin{align}
\hat{\theta}_{t,k} &= \tan^{-1}\left( \frac{\hat{\omega}_{y,k} }{\hat{\omega}_{x,k}} \right) \label{vartheta} \\
\hat{\phi}_{t,k} &= \sin^{-1}\left( \sqrt{\left( \frac{\nu}{2\pi d}\hat{\omega}_{x,k}\right)^2 + \left( \frac{\nu}{2\pi d}\hat{\omega}_{y,k}\right)^2}  \right) \label{varphi}.
\end{align}


\subsection{Path-loss Estimation}\label{section4c}
The path-loss $\bbe$ is merged into the factor matrices. Unlike DOA/DOD estimation which is insensitive to the column scaling ambiguity, the estimation of $\bbe$ is seriously affected by such an ambiguity. 

To estimate $\bbe$, let us find the explicit expression for estimating $\bbe$. 
Note that the column permutation in $\{\hat\B_r,\hat\C_x,\hat\C_y\}$ is not an issue, let us ignore it and consider only the scaling ambiguity to simplify the analysis. 
The reconstruction of $\Y = \H\S$ from the estimated factor matrices is given by
\begin{align}\label{Yhat}
	\Y = \H\S
	&= \hat{\B}_r\(\hat{\C}_y\odot\hat{\C}_x\)^H \notag\\
	&= \A_r\bXi_r\bXi_y^*\bXi_x^*\(\A_y\odot\A_x\)^H\(\S_y\otimes\S_x\) \notag\\
	&= \A_r\bXi_r\bXi_y^*\bXi_x^*\(\A_y\odot\A_x\)^H\S.
\end{align}
Since $\H\S = \A_r\diag(\bbe)\(\A_y\odot\A_x\)^H\S$, by comparing it with \eqref{Yhat}, we find
\begin{align}\label{bbe}
	\diag\big(\bbe\big) = \bXi_r\bXi_y^*\bXi_x^*
\end{align}
where $\bXi_r = \diag([\xi_{r,1},\cdots,\xi_{r,K}])$ is the scaling ambiguity matrix corresponding to $\A_r$ with its $i$th diagonal entry being
\begin{align}\label{bbe2}
	\beta_k = \xi_{r,k}\xi_{y,k}^*\xi_{x,k}^*,~\forall k=1,\cdots,K. 
\end{align}
In the above, the scaling ambiguity corresponding to $\A_r$ is estimated as
\begin{align}\label{ss1}
	\hat\xi_{r,k} = \hat{\a}_{r,k}^\dagger\hat{\b}_{r,k}
\end{align}
where $\hat{\a}_{r,k}$ is constructed using $\hat\omega_{x,k}$. The only unknown in \eqref{bbe2} is  $\hat\xi_{y,k}^*\hat\xi_{x,k}^*$. We propose an efficient forward-backward average method to calculate it which only involves element-wise multiplication/division.

\subsubsection{Forward Way}
First, let us consider the forward way. Let
\begin{align*}
 \overline{e}_{i,k} &= \overline{\s}_{i,1}^H\a_{i,k} \\
 \overline{f}_{i,k} &= \overline{\s}_{i,2}^H\a_{i,k}, \forall i\in\{x,y\}
\end{align*}
be the $N_i/2$ and $(N_i/2-1)$ elements in $[\overline\C_i]_{:,k}$, respectively.

In the presence of scaling ambiguity, $\hat{\overline\C}_i$ is expressed as
\begin{align*}
 \hat{\overline\C}_i = \overline\C_i{\bXi}_i
\end{align*}
where ${\bXi}_i = \diag([\xi_{i,1},\cdots,\xi_{i,K}])$ contains $K$ scaling ambiguities with $\xi_{i,k}$ standing for the ambiguity between $[\hat{\overline\C}_i]_{:,k}$ and $[{\overline\C}_i]_{:,k}$.
The estimates of $\overline{e}_{i,k}$ and $\overline{f}_{i,k}$ now become
\begin{align}
 \hat{\overline{e}}_{i,k} &= \overline{\s}_{i,1}^H\a_{i,k}\xi_{i,k} \label{ana1}\\
 \hat{\overline{f}}_{i,k} &= \overline{\s}_{i,2}^H\a_{i,k}\xi_{i,k}
\end{align}
and
\begin{align}
 \hat{\overline{e}}_{i,k} = \hat{\overline{f}}_{i,k} + [\overline\s_{i,1}]_{M_i}^*[\a_{i,k}]_{M_i}\hat\xi_{i,k}
\end{align}
It follows that
\begin{align}\label{xiforward}
 \hat\xi_{i,k} 
 = \frac{\hat{\overline{e}}_{i,k} - \hat{\overline{f}}_{i,k}}{[\a_{i,k}]_{M_i} [\overline\s_{i,1}]_{M_i}^*}
 = \frac{\hat{\overline{e}}_{i,k} - \hat{\overline{f}}_{i,k}}{e^{j(M_i-1)\hat\omega_{i,k}} [\overline\s_{i,1}]_{M_i}^*}
\end{align}
where $[\a_{i,k}]_{M_i}$ is replaced by its estimate $e^{j(M_i-1)\hat\omega_{i,k}}$.
Then we have
\begin{align}\label{ss2}
&\hat\xi_{y,k}^*\hat\xi_{x,k}^*  \notag\\
&= \bigg(\frac{\hat{\overline{e}}_{y,k} - \hat{\overline{f}}_{y,k}}{e^{j(M_y-1)\hat\omega_{y,k}} [\overline\s_{y,1}]_{M_y}^*} \bigg)^*
\bigg(\frac{\hat{\overline{e}}_{x,k} - \hat{\overline{f}}_{x,k}}{e^{j(M_x-1)\hat\omega_{x,k}} [\overline\s_{x,1}]_{M_x}^*}\bigg)^* \notag\\
&= \bigg(\frac{\hat{\overline{e}}_{y,k} - \hat{\overline{f}}_{y,k}}{e^{j(M_y-1)\hat\omega_{y,k}} } \bigg)^*
\bigg(\frac{\hat{\overline{e}}_{x,k} - \hat{\overline{f}}_{x,k}}{e^{j(M_x-1)\hat\omega_{x,k}} }\bigg)^*[\overline\s_{y,1}]_{M_y}[\overline\s_{x,1}]_{M_x}
\notag\\
&= \bigg(\frac{\hat{\overline{e}}_{y,k} - \hat{\overline{f}}_{y,k}}{e^{j(M_y-1)\hat\omega_{y,k}} } \bigg)^*
\bigg(\frac{\hat{\overline{e}}_{x,k} - \hat{\overline{f}}_{x,k}}{e^{j(M_x-1)\hat\omega_{x,k}} }\bigg)^*
\end{align}
where the last equality is due to \eqref{ss}.

\subsubsection{Backward Way}
To maximize the information usage, for example, we may also calculate $\xi_{y,k}^*\xi_{x,k}^*$ from the $(N_i-1)$th and $N_i$th rows of $[\underline\C_i]_{:,k}$. The derivations are mostly the same as the forward way. The only difference is that $\underline{\s}_{i,k}$ is conjugate flipped from $\overline{\s}_{i,k}$. We have the following relationship
\begin{align*}
\underline{e}_{i,k} &= [\a_{i,k}]_{2}\underline{f}_{i,k} + [\underline\s_{i,1}]_{1}[\a_{i,k}]_{1} \notag\\
&= \underline{f}_{i,k}e^{j\omega_{i,k}} + [\underline\s_{i,1}]_{M_i}
\end{align*}
where $\underline{e}_{i,k} = \underline{\s}_{i,1}^H\a_{i,k}$ and $\underline{f}_{i,k} = \underline{\s}_{i,2}^H\a_{i,k}$.
Following the analysis in \eqref{ana1}--\eqref{xiforward} yields
\begin{align*}
\hat\xi_{i,k} = \frac{\hat{\underline{e}}_{i,k} - \hat{\underline{f}}_{i,k}e^{j\hat\omega_{i,k}}}{[\underline\s_{i,1}]_{M_i}}.
\end{align*}
Then based on \eqref{ss}, we have
\begin{align}\label{ss4}
\hat\xi_{y,k}^*\hat\xi_{x,k}^* &= \(\hat{\underline{e}}_{y,k} - \hat{\underline{f}}_{y,k}e^{j\hat\omega_{y,k}}\)^*\(\hat{\underline{e}}_{x,k} - \hat{\underline{f}}_{x,k}e^{j\hat\omega_{x,k}}\)^*.
\end{align}
Finally, we substitute the average of \eqref{ss2} and \eqref{ss4} into \eqref{bbe2} for final path-loss estimation.

Our method contains two main procedures: tensor decomposition and multipath parameter estimation. Both of them exploit the rotational invariance property which exists in the Vandermonde manifold matrices. Because of this reason, we name our method as \textit{rotationally invariant channel estimation (RICE)} algorithm. Its detailed steps are summarized in Algorithm \ref{algorithm2}.

\subsection{Identifiability Analysis}
The last remaining question is identifiability, i.e., how many paths we can handle given the measurements in \eqref{y}.
Recall that the dimension of $\Z$ in \eqref{Z} is a function of $P_r$ and $Q_r$. Since $N_x$ and $N_y$ are fixed, by tuning $P_r$ and $Q_r$, we are able to find an optimal pair of $\{P_r,Q_r\}$ such that the number of paths that our method is capable to cope with is maximized. Based on Theorem 1 \cite{smoothESPRIT}, we have the following result.
\begin{theorem}\label{theorem}
	Assume that the DOAs and DODs in different paths are not identical, i.e., $\theta_{r,i}\neq\theta_{r,j},\theta_{t,i}\neq\theta_{t,j},\phi_{t,i}\neq\phi_{t,j}, \forall i\neq j$, and all the path-losses are jointly drawn from an absolutely continuous distribution. Then, given the measurements $\Y$, all the multipath parameters are uniquely identifiable with probability one if
	\begin{align}
	K \leq \min\Big(  (P_r-1)N_x, ~Q_rN_y  \Big)
	\end{align}
	where $P_r$ and $Q_r$ are chosen from
	\begin{align}\label{max}
	\max_{P_r,Q_r}~ &\min\Big((P_r-1)N_x,~ Q_rN_y\Big).
	\end{align}
\end{theorem}

The proof of Theorem \ref{theorem} is constructive, following the steps of Algorithm 1, meaning that this bound is achievable. In practice, once $M_r$ and $N$ are chosen, we first find the optimal $\{P_r,Q_r\}$ by solving \eqref{max} and then cache them in the system to guarantee the identifiability. 
We note that the minimum $N$ is 16. If we choose $M_r=N_x=N_y=4$, our method can uniquely identify up to eight paths, while the standard array processing methods can only handle three paths.

\begin{algorithm}[t]
	\caption{RICE}
	\begin{algorithmic}[1]\vspace{.3em}
		\State Determine the optimal $P_r$ and $Q_r$ from Theorem \ref{theorem}.
		\State Use Algorithm 1 to identify $\hat{\B}_r,\hat{\C}_x,\hat{\C}_y$.
		\State Estimate $\omega_{r,k}$ via \eqref{omegar}, $\omega_{x,k}$ and $\omega_{y,k}$ from \eqref{omegax} and \eqref{omegay}, respectively.
		\State Compute $\hat\bXi_r$ via \eqref{ss1}, average \eqref{ss2} and \eqref{ss4} for $\hat\bXi_y^*\hat\bXi_x^*$, and then estimate $\hat\bbe$ via \eqref{bbe}.
		\State Recover the channel matrix from
		$\{\hat\omega_{r,k},\hat\omega_{x,k},\hat\omega_{y,k},\hat\beta_k\}_{k=1}^K$.
	\end{algorithmic}\label{algorithm2}
\end{algorithm}

\subsection{Complexity Analysis}
The computational complexity for the proposed method mainly lies in Algorithm 1, where the SVD of $\tilde{\Z}$ in Step 2 costs about $\cO(P_r^2N_x^2Q_rN_y)$ flops. Since $P_r+Q_r=M_r+1$ and $M_r$ is usually no more than four according to nowadays technology, we have $P_r^2Q_r\approx 18\approx M_r^2$. On the other hand, because of the fact $N = N_xN_y$, by setting $N_x=N_y=\sqrt{N}$, the complexity of Step 2 becomes $\cO(M_r^2N ^{1.5})$. 
After the factor matrices are obtained, the estimation of DOAs, DODs and path-losses are very simple.
The DOA estimation in \eqref{doa} requires about $\cO(M_rK)$ flops. The estimation of azimuth and elevation angles in \eqref{omegax}--\eqref{varphi} costs about $\cO(LK)$ flops where $L=N^{1/4}$. In many cases $L=2$ is enough to achieve satisfactory performance. Thus, $\cO(LK)\approx\cO(K)$. 
The calculation of $\bbe$ costs $\cO(M_rK+K)$ flops.
The overall complexity of the proposed method is $\cO(M_r^2N ^{1.5}+2M_rK+3K)$, which is quite low compared to the sparse regression methods such as orthogonal pursuit (OMP) that requires $\cO(M_rNK2^{21})$ flops when DOD and DOA are quantized with $7$ bits.

\vspace{.5em}
\begin{remark}	
	The main advantages of the proposed method are its low-complexity and identifiability guarantees.	As far as we know, there are very few uniqueness results available for MIMO channel estimation in the case of $N\ll M_t$, i.e., when there are less training samples than the number of transmit antennas. And there is a serious lack of efficient and reliable channel estimation algorithms to handle such difficult cases especially when $M_r$ is small and $K$ is relatively large. 
	In \cite{qc2018}, we considered similar cases for a special type of MIMO systems with dual-polarized antennas, where we analyzed the identifiability and proposed algorithms for channel estimation. Unfortunately, the results in \cite{qc2018} are only valid for dual-polarized MIMO and hence, cannot be applied here. Therefore, the results in this paper are much more general---and also timely and meaningful as 5G system trials are beginning to roll out. It is worth highlighting that the proposed method can be generalized to the dual-polarized systems and similar closed-form solutions for multipath parameter estimation can be derived in a straightforward way.
\end{remark}

\section{Exploiting full knowledge of ${\bf S}$}
	The RICE algorithm is based on the spatial structure of the training sequence for channel estimation but not the values. This means that only partial information of the training sequence has been used in RICE. 
	Because of this reason, the complexity of RICE is maintained at a very low level, but at the expense of losing resolution ability in DOD estimation.
	In this section, we show that by fully utilizing the information in the training sequence, DODs can be estimated through a root-finding technique, and the performance of RICE can be further improved with  a moderate increase of complexity. Toward this end, a  joint \textbf{RICE} and \textbf{R}oot-finding approach is developed. We name the new algorithm \textit{RICER}. 

	In RICER, the estimation of DOA is the same as RICE, i.e., \eqref{doa}. 
	The difference is in the estimation of DOD and path-loss. 
	First, let us consider the estimation of $\omega_{x,k}$ from $\c_{x,k}=\S_x^H\a_{x,k}$.
	Before we proceed, we claim that with the following Corollary, $\omega_{x,k}$ is identifiable from $\c_{x,k}$.
	\begin{corollary}\label{cor:QA}\cite{qc2018}
		Assume that $M,N\geq 2$, and that $\A\in\bC^{M\times F}$ is Vandermonde with distinct generators, i.e., $\omega_i\neq \omega_j$ for $i\neq j$. Then, $\omega$ can be uniquely identified almost surely from the system $\C=\Q^H\A(\omega)\diag(\bxi)$, where $\bxi$ stands for column scaling and the elements in $\Q\in\bC^{M\times N}$ are jointly drawn from an absolutely continuous distribution.
	\end{corollary}

    We note that in the absence of noise,
    \begin{align}\label{span1}
    \mathrm{span}(\c_{x,k})  = \mathrm{span}(\S_x^H\a_{x,k}).
    \end{align}
    The above is interesting, indicating that although $\S_x^H$ is fat, its null space does not contain any Vandermonde vector.
   The above also says that the orthogonal complement of $\c_{x,k}$ is orthogonal to $\S_x^H\a_{x,k}$. Thus, we have
    \begin{align}\label{root0}
    \left\|\(\I_{M_x} - \frac{\c_{x,k}\c_{x,k}^H}{\|\c_{x,k}\|_2^2}\)\S_x^H\a_{x,k}\right\|_2^2 = 0.
    \end{align}
    Define
    \begin{align}
    \P_{x,k}^\perp = \S_x\(\I_{M_x} - \frac{\c_{x,k}\c_{x,k}^H}{\|\c_{x,k}\|_2^2}\)\S_x^H.
    \end{align}
    Eq. \eqref{root0} is then equivalent to
    \begin{align}\label{root1}
     \a_{x,k}^H\P_{x,k}^\perp\a_{x,k} = 0.
    \end{align}
    In the noisy case, the equality in \eqref{root1} approximately holds true. We can then estimate $\omega_{x,k}$ via 
    \begin{align}\label{music}
    \min_{\omega_{x,k}} \a_{x,k}^H\P_{x,k}^\perp\a_{x,k}
    \end{align}
    which is similar to the cost of the classical MUSIC algorithm \cite{schmidt1986multiple}. Thus, we may search the phase from $-\pi$ to $\pi$ and report the one that minimizes the objective function in \eqref{music}. The drawback is the complexity caused by the 1-D angular search which is approximately $\cO(M_x^2D)$ flops for each $\omega_{x,k}$, where $D$ is the number of bins dividing $[-\pi,\pi]$. We may also derive a gradient descent method to handle \eqref{music} \cite{qc2018}. But due to the non-linearity and non-convexity in \eqref{music}, optimizing the phase requires careful initialization which is not easy to be acquired in an efficient way.
    
    Here, we employ a root-finding technique to estimate \eqref{omegax}. Since $\a_{x,k}$ is Vandermonde, we further express \eqref{root1} as a polynomial:
	\begin{align}\label{rm:coef}
	c_0 + c_1z^{-1} + \cdots + c_{2M_x-1}z^{-(2M_x-1)} = 0
	\end{align}
	where 
	\begin{align}
		c_i = 
		\begin{cases}
			\sum_{m=1}^{i} [\P_{x,k}^\perp]_{M_x-i+m,m+1},~ i=0,\cdots,M_x\\
			\sum_{m=1}^{2M_x-i} [\P_{x,k}^\perp]_{m,i+m-M_x},~ i=M_x+1,\cdots,2M_x-1.
		\end{cases}
	\end{align}
	
	There are totally $(2M_x-2)$ roots after solving \eqref{rm:coef}. The symmetric property of $\P_{x,k}^\perp$ implies that half of the roots are inside the unit circle while another half are outside, and they appear in conjugate-reciprocal pairs. In other words, the outer roots are inverses of the inner roots. We are only interested in the inner roots, i.e., those inside the unit circle. Let us denote them as $\mathcal{Z} = \{z_i\mid|z_i|\leq 1, i=1,\cdots,M_x-1\}$. The next problem is to judiciously select one from $\mathcal{Z}$ for estimating $\omega_{x,k}$. 
	It is well motivated from the philosophy of root-MUSIC \cite{rao1989performance} that one may find a root $z$  from $\mathcal{Z}$ that is closest to the unit circle.
	However, when SNR is low or sample size is small, the signal subspace $\c_{x,k}$ might be heavily corrupted by a portion of noise subspace which causes the subspace leakage problem \cite{shaghaghi2015subspace}. Once this happens, the root-MUSIC rule---selecting the root that is closest to the unit circle---is problematic; some irrelevant roots from the orthogonal complement of $\c_{x,k}$ may be much closer to the unit circle than the true root. 
	Such phenomenon happens oftentimes when we estimate $\omega_{x,k}$ since $\hat\c_{x,k}$ is noisy and the available degrees-of-freedom are only $N_x\approx4$. 
	Provided that the wrong root is selected, we can never reconstruct the channel correctly.
	Therefore, it is crucial to design a robust rule for final root determination. 
	
	One way to help alleviate the subspace leakage issue is as follows. Note that the RICE method relies more on the Vandermonde structure of $\v_{x,k}$ but not the subspace $\c_{x,k}$, so RICE is robust for subspace leakage. As a result, we can use the estimate of $\omega_{x,k}$ from RICE for assistance.
	Specifically, let us calculate the phases in $z_i$ as $\psi_i=\angle(z_i),\forall i=1,\cdots,M_x-1$. Let $\tilde\omega_{x,k}$ denote the estimate of $\omega_{x,k}$ from RICE, i.e., \eqref{omegax}. We select one from $\{\psi_i\}$ that is closest to $\tilde\omega_{x,k}$ as the final estimate of $\omega_{x,k}$.

	Following the same way, we can calculate $\hat\omega_{y,k}$ from $\c_{y,k}$. Finally, we update the path-loss as
	\begin{align}\label{knownSbeta}
		\hat\bbe = \(\big(\S^T(\hat\A_y\odot\hat\A_x\big)^*) \odot \hat\A_r\)^\dagger \vec(\Y)
	\end{align}
	where $\hat\A_i$ is constructed from $\{\hat\omega_{i,1},\cdots,\hat\omega_{i,K}\},~i=r,x,y$ which are obtained from RICER. The detained steps for RICER are provided in Algorithm \ref{algorithm3}.

	The identifiability of RICE and RICER is basically the same since both methods rely on Algorithm 1 for tensor decomposition. Their main difference lies in the estimation of DOD and path-loss. 
	RICER uses the aid of RICE for determining $\omega_{x,k}$ and $\omega_{y,k}$. Thus, its complexity is higher than RICE. We have $2K$ phases in total. The related complexity for solving $2K$ polynomials in \eqref{rm:coef} is $\cO\(8K(M_x^2\log(M_x)+M_y^2\log(M_y))\)$ flops. The complexity for updating the path-loss using \eqref{knownSbeta} is $\cO(M_rM_xM_yNK)$ flops. The total complexity for RICER is $\cO\big(M_r^2N^{1.5} + 8K(M_x^2\log(M_x)+M_y^2\log(M_y)) + M_rM_xM_yNK\big)$ flops which is much higher than the complexity of RICE. In the next section, we will see that by paying this additional complexity, RICER achieves better performance than RICE.

\begin{algorithm}
	\caption{RICER}
	\begin{algorithmic}[1]\vspace{.3em}
		\State Determine the optimal $P_r$ and $Q_r$ from Theorem \ref{theorem}.
		\State Use Algorithm 1 to identify $\hat{\B}_r,\hat{\C}_x,\hat{\C}_y$.
		\While{$k=1,\cdots,K$}
		\State Estimate $\omega_{r,k}$ via \eqref{omegar}; use the root finding algorithm described in Section V to estimate $\omega_{x,k}$ and $\omega_{y,k}$ in $\hat\c_{x,k}$ and $\hat\c_{y,k}$, respectively.
		\EndWhile
		\State Reconstruct $\A_r$, $\A_x$ and $\A_y$ from $\hat\omega_{r,k}$, $\hat\omega_{x,k}$ and $\hat\omega_{y,k}$, respectively, and calculate the path-losses using \eqref{knownSbeta}.
		\State Recover the channel matrix from $\{\hat\omega_{r,k},\hat\omega_{x,k},\hat\omega_{y,k},\hat\beta_k\}_{k=1}^K$.
	\end{algorithmic}\label{algorithm3}
\end{algorithm}

\subsection{Special Case: $M_r=1$}

At this point, the reader might wonder whether the proposed framework can work in the case where only one antenna is available at the mobile end. The answer is affirmative.
Let us first take a look at the signal model with a single receive antenna
\begin{align}
\tilde\y = \(\S_y^H\A_y \odot \S_x^H\A_x\)\bbe
\end{align}
which can be reshaped into a matrix as
\begin{align}\label{tildeY}
\tilde{\Y} = \S_x^H\A_x\diag(\bbe)\(\S_y^H\A_y\)^T \in \bC^{N_x\times N_y}
\end{align}
We see that the tensor structure is no longer available in the received signal, and therefore uniqueness of the factor matrices seems to fail too. Since the RICE and RICER algorithms require the identification of $\S_x^H\A_x$ and $\S_y^H\A_y$ before performing parameter estimation, both of them will not work in the single antenna case. However, some further reflection shows that the RICER method can be modified for channel estimation even with a single receive antenna. 

Note that when the training signals are orthogonal, channel estimation from \eqref{tildeY} is indeed a 2-D harmonic retrieval problem, which is not our interest. Therefore, we only consider ``tall'' $\S_x$ and $\S_y$, i.e., $N_x<M_x$ and $N_y<M_y$. 
Let $\tilde{\U}_s\in\bC^{M_x\times K}$ be the signal subspace of $\tilde{\Y}$. Similar to \eqref{span1}, we have
\begin{align}
\mathrm{span}(\S_x^H\A_x) = \mathrm{span}(\tilde{\U}_s).
\end{align}
Then the following equation holds
\begin{align}
\a_{x,k}^H\tilde{\P}_{x}^\perp\a_{x,k} = 0
\end{align}
where 
$$\tilde{\P}_{x}^\perp = \S_x\(\I_{M_x} - \tilde{\U}_{x}\tilde{\U}_{x}^H\)\S_x^H.$$
Owing to the Vandermonde structure, we can also employ the root-finding technique to estimate $\omega_{x,k},\forall k=1,\cdots,K$ by solving \eqref{rm:coef}, where $\P_{x,k}^\perp$ is replaced by $\tilde{\P}_{x}^\perp$. Then we pick the top $K$ roots inside of the unit circle and estimate $\omega_{x,k}$ as the phase of the $k$th root. After that we use the estimates $\{\hat\omega_{x,k}\}$ to construct $\hat\A_x$ and calculate
\begin{align}
\tilde{\C}_y = \( \(\S_x^H\hat\A_x\)^\dagger\tilde{\Y} \)^T.
\end{align}
Following \eqref{span1}--\eqref{rm:coef}, we can find the estimates of $\{\hat\omega_{y,k}\}$.
Finally, estimate the path-losses as $\big(\S_y^H\hat\A_y\odot\S_x^H\hat\A_x\big)^\dagger\tilde\y$. 

Note that the channel parameters can be identified via the above procedures if $\A_x$ and $\A_y$ are full column rank and $K<\min(N_x,N_y)$.

\section{Simulations}
In the simulation, we assume that the multipath propagation gains are Rician distributed. 
All the results are averaged over 500 Monte-Carlo trials using a computer with 3.7 GHz Intel Core i7-8700 and 32 GB RAM.
The normalized mean square error (NMSE) of channel estimates is computed from
$
	\mathrm{NMSE} = \frac{1}{500}\sum_{i=1}^{500}\|\hat{\H}_{i}-\H_i\|_F^2/\|\H_i\|_F^2
$
where $\hat{\H}_{i}$ denotes the channel that is reconstructed from the estimated multipath parameters from the $i$th Monte-Carlo trial.
The estimation of $K$ is beyond the scope of this paper. So in the simulations, to be fair, we assume that $K$ is known to all the algorithms.

Given the model in \eqref{y}, array processing methods fail to work when $K\geq M_r$. The existing methods which are qualified to handle large $K$ might be the class of CS based methods \cite{jarvis,jomp,lee2016channel}. We choose the OMP method for performance comparison since it is hyper-parameter free and computationally efficient.
To implement OMP, we quantize $\theta_r$, $\theta_t$ and $\phi_t$ using 7 bits, so the resulting dictionary has size $4N\times 2^{21}$.
We only consider ``tall'' $\S$, i.e., there are less samples than transmit antennas. Hence, the LS technique does not work. 
We are interested in how well the new methods perform with a `tall' $\S$ compared to the LS estimate with an orthogonal square $\S$. The best achievable NMSE of the LS channel estimate from orthogonal training is $10^{-\mathrm{SNR}/10}$, where SNR is in dB. We include this value as a performance benchmark.

\begin{figure}[t]
	\begin{center}
		\subfigure[$M_r=3,K=4$]{\label{fig:location1} \includegraphics[width=1\linewidth]{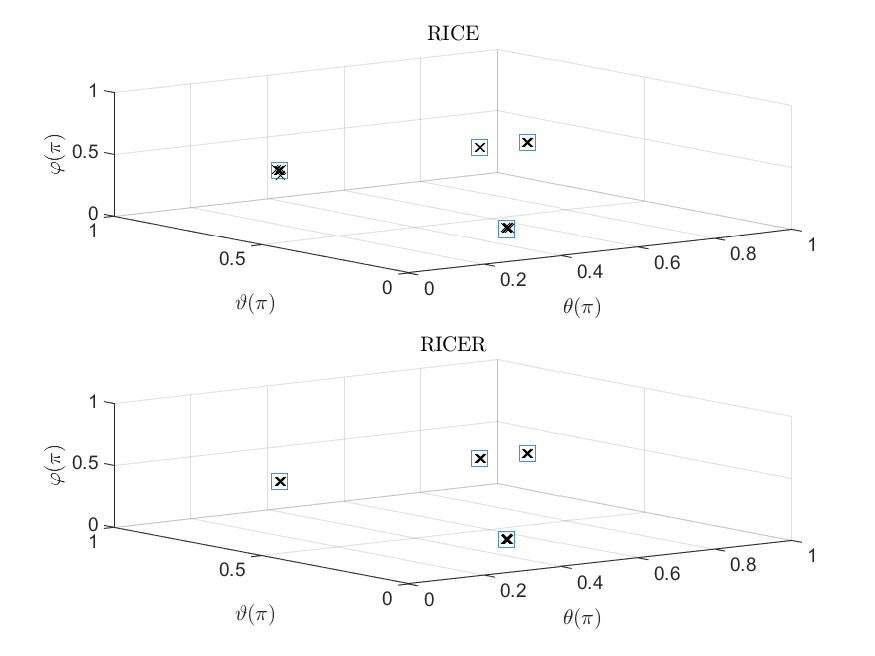}} 
		\subfigure[$M_r=4,K=8$]{\label{fig:location2} \includegraphics[width=1\linewidth]{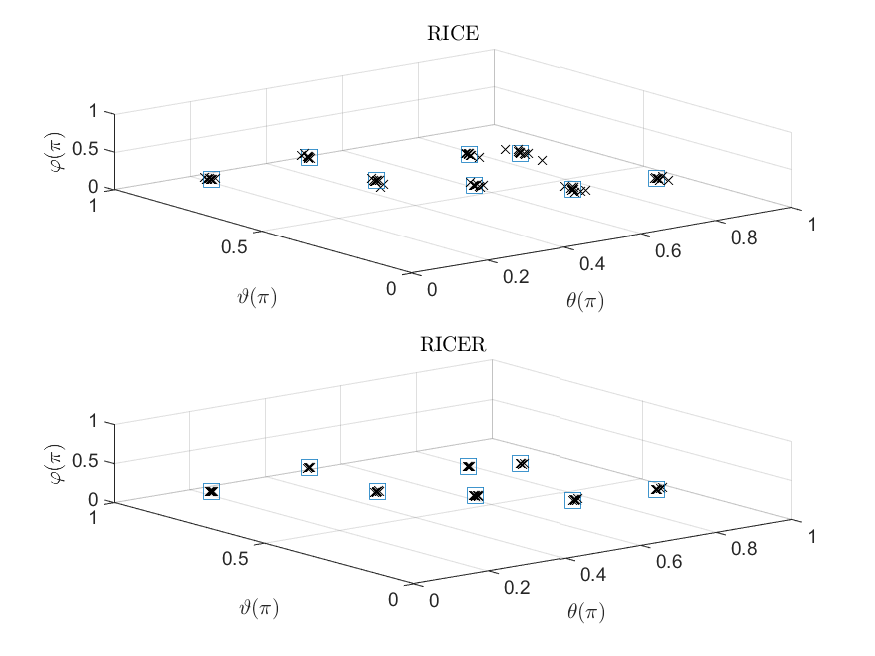}}
	\end{center}
	\caption{Verification of identifiability, where $\{\theta_r,\theta_{t},\phi_{t}\}$ of each path are plotted to localize the scatterers. In the figures, $\small\Box$ denotes the true location of scatterers and $\times$ denotes the estimates. The number of paths is set to be the maximum $K$ from Theorem 2.}\label{fig:identifiability}
\end{figure}

In the beginning, we examine the identifiability of our methods. Let us consider the following parameter setting: $M_r=3$, $M_x=M_y=10$, $N_x=N_y=4$ and SNR $=20$ dB. We set the number of paths $K$ as the maximum number of identifiable paths calculated based on Theorem 2. Under this setting, we have $K=4$. The phases $\{\omega_{r,k},\omega_{t,K},\omega_{t,K}\}$ are chosen as
\begin{align*}
\begin{bmatrix}
\bom_r^T \\ \bom_x^T \\ \bom_y^T
\end{bmatrix} = \pi \times
\begin{bmatrix}
0.8  & 0.57 & 0.1  & 0.33 \\
0.36 & 0.7 & 0.53 & 0.2 \\
0.8  & 0.33 & 0.57 & 0.1 
\end{bmatrix}.
\end{align*}
Fig. \ref{fig:location1} plots the locations of each scatter based on their DOAs and DODs. It shows that RICE and RICER are able to resolve all the paths. Next we choose $M_r=4$ and SNR $=30$ dB. According to Theorem 2, our methods can deal with $K=8$ paths in theory. To verify this, we set
\begin{align*}
\begin{bmatrix}
\bom_r^T \\ \bom_x^T \\ \bom_y^T
\end{bmatrix} = \pi\times
\begin{bmatrix}
0.8  \!&\!\! 0.4  \!\!&\!\! 0.3  \!\!&\!\! 0.1  \!\!&\!\! 0.5  \!&\!\! 0.2  \!\!&\!\! 0.7  \!\!&\!\! 0.6 \\
0.34  \!&\!\! 0.49  \!\!&\!\! 0.41  \!\!&\!\! 0.27  \!\!&\!\! 0.56  \!\!&\!\! 0.7  \!\!&\!\! 0.2  \!\!&\!\! 0.63 \\
0.2  \!&\!\! 0.3  \!\!&\!\! 0.5  \!\!&\!\! 0.8  \!\!&\!\! 0.1  \!&\!\! 0.6  \!\!&\!\! 0.7  \!\!&\!\! 0.4
\end{bmatrix}.
\end{align*}
The results are shown in Fig. \ref{fig:location2}, where our algorithms still work well. However, some angle estimates of RICE slightly disperse around the true angles, while those of RICER are more concentrated. 

\begin{figure}
	\centering
	\includegraphics[width=0.9\linewidth]{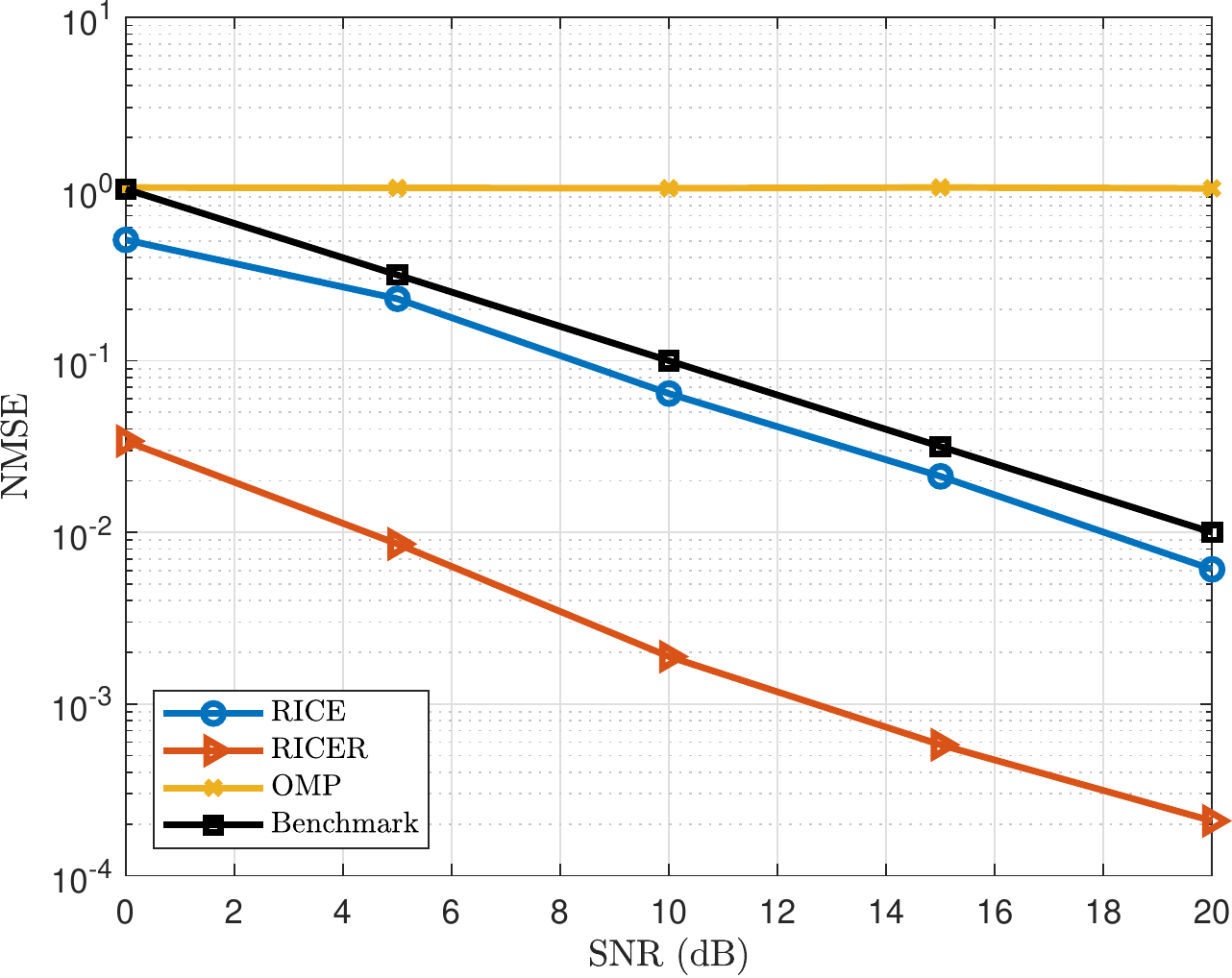}
	\caption{NMSE versus SNR.}
	\label{fig:nmsevssnrfixedk}
\end{figure}

\begin{figure}
	\centering
	\includegraphics[width=0.9\linewidth]{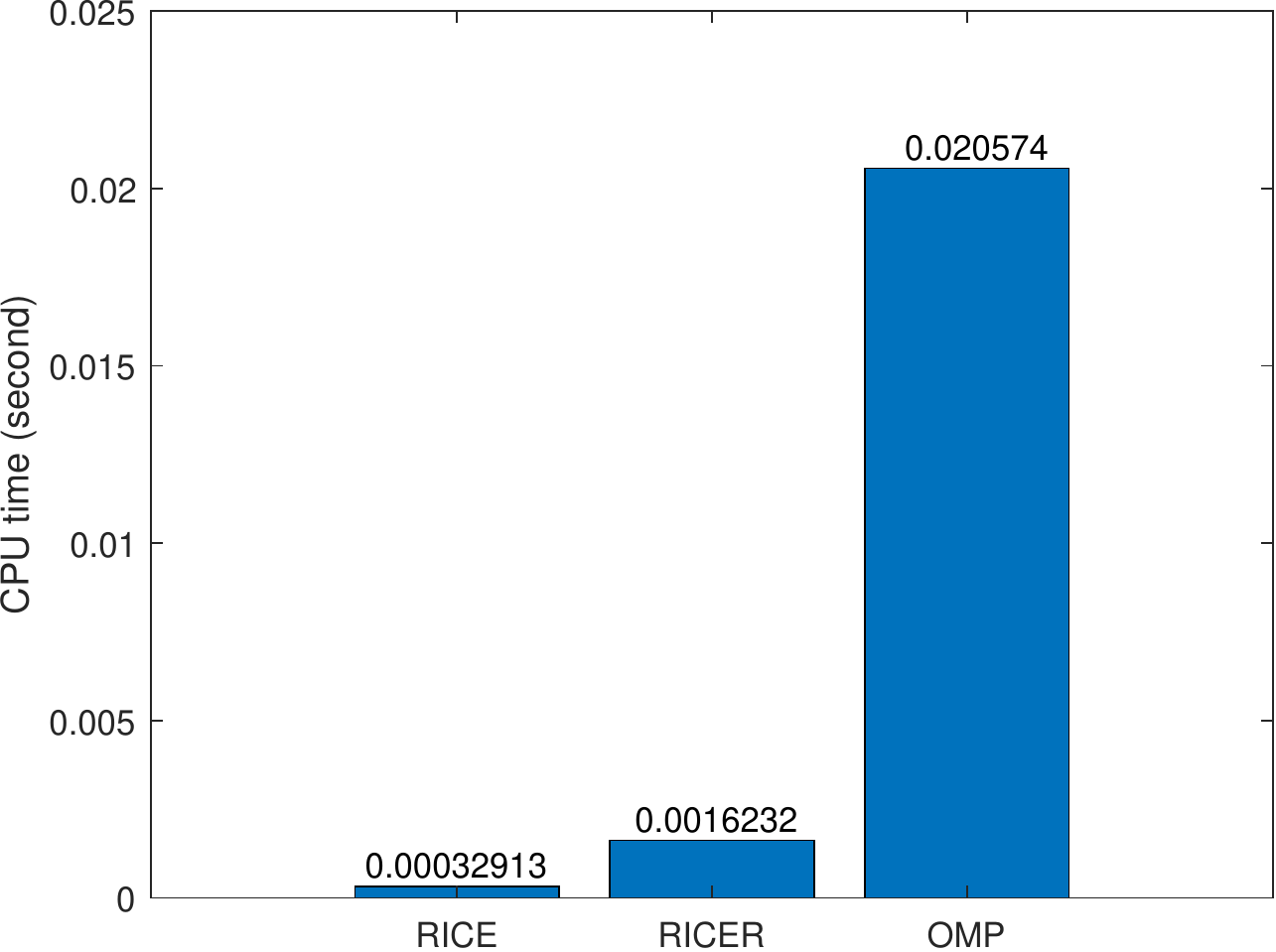}
	\caption{Complexity comparison.}
	\label{fig:cputime}
\end{figure}

Now let us study the NMSE performance of the proposed methods. We first compare the NMSE performance by varying SNR from 0 to 20 dB. We set $M_r=4$, $M_x=M_y=10$,  $K=4$ and $N_x=N_y=4$. 
Simulation results are provided in Fig. \ref{fig:nmsevssnrfixedk}, where both RICE and RICER outperform the OMP and benchmark throughout the range of SNRs considered. Note that the benchmark here uses a training signal of length 100 (vs. 16 for RICE and RICER), but does not exploit the DOA-DOD path parametrization -- this is why RICE can beat this benchmark. Also note that OMP, which leverages the DOA-DOD path parametrization, does not work well due to the lack of enough samples and the coherence in the dictionary. RICER has better accuracy than RICE but it is shown in Fig. \ref{fig:cputime} that this is at the expense of paying four times more complexity. The additional calculation time is caused by the root finding procedure for DOD estimation and LS for path-loss estimation. Notably, RICE is 62 times faster than OMP and RICER is 12 times faster.

\begin{figure}
	\centering
	\includegraphics[width=0.9\linewidth]{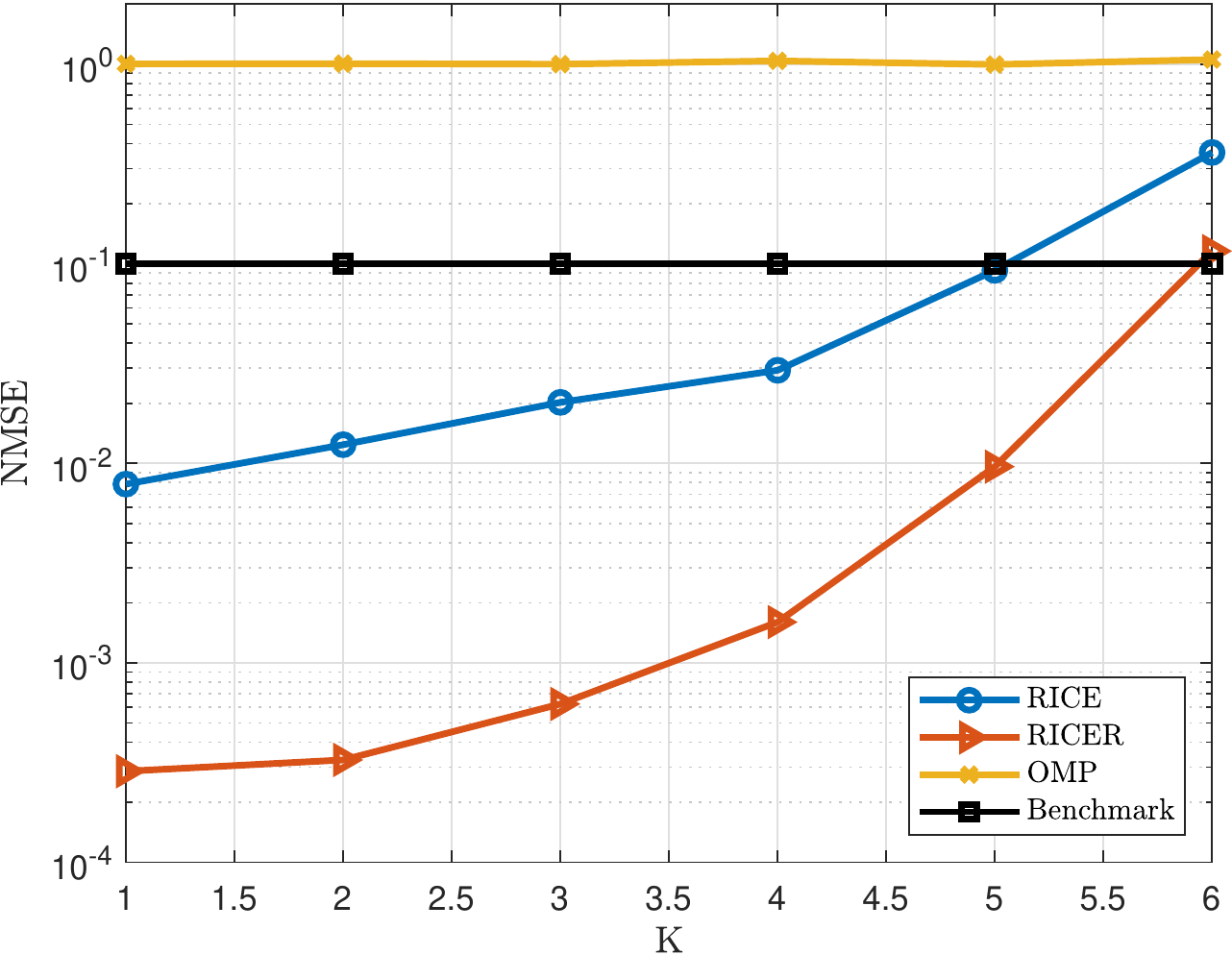}
	\caption{NMSE versus $K$.}
	\label{fig:nmsevsk}
\end{figure}

Similar results can also be found in Fig. \ref{fig:nmsevsk}, where the number of paths varies from 1 to 6 and SNR is fixed at 10 dB. $\{\omega_{r,k}\}$ are generated by uniformly dividing the range $(0.4\pi, 1.6\pi)$ into $K$ intervals. $\{\omega_{x,k}\}$ and $\{\omega_{y,k}\}$ are generated in the same way but from the range $(0.4\pi,1.8\pi)$ and $(0.2\pi, 1.6\pi)$, respectively.
Again, OMP does not work, likely owing to the coherence of the `flat'  dictionary matrix whose dimension is $64\times2^{21}$. RICE and RICER offer satisfactory performance for small $K$. 
However, when $K$ exceeds 5, RICE becomes a bit inferior to the benchmark but still acceptable for such a difficult setting---recovering 5 paths from 16 training samples, which indicates that using full knowledge of $\S$ helps in achieving better estimation accuracy but at the expense of high complexity.
Also we point out that according to Theorem 2, with the parameter settings of this example, our methods can resolve up to $K=6$ distinct paths. Therefore, the performance loss of RICE and RICER is due to the fact that the number of paths reaches the upper-bound that they can handle.

\begin{figure}
	\centering
	\includegraphics[width=0.9\linewidth]{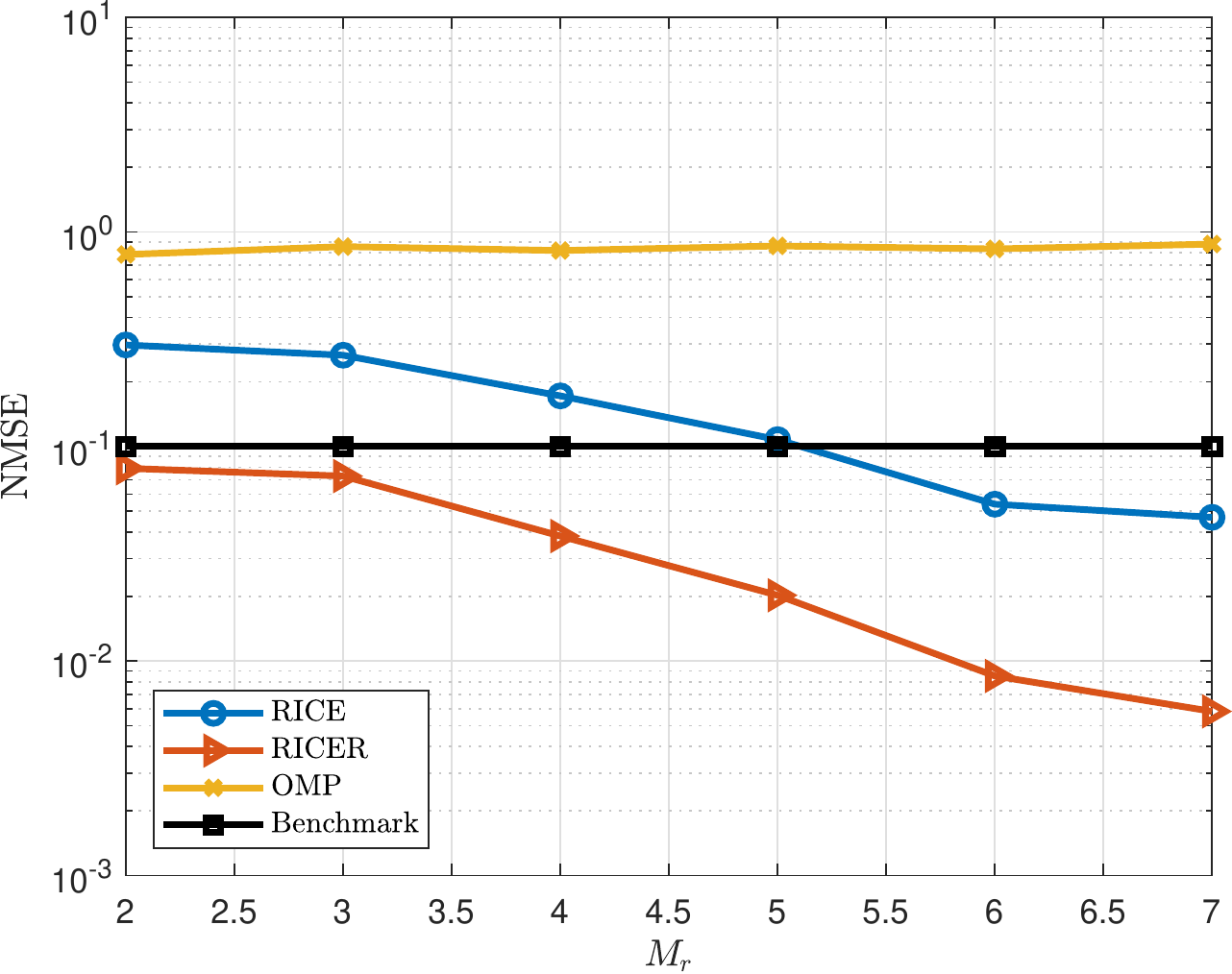}
	\caption{NMSE versus $M_r$.}
	\label{fig:nmsevsmr}
\end{figure}

Next we evaluate the performance as a function of $M_r$. We set $M_x=M_y=10$, $N_x=N_y=6$, SNR $=10$ dB, and vary $M_r$ from 2 to 7. At the same time, we increase the number of paths $K$ for each $M_r$ in the range $\{2, 3, 4, 5, 6, 7\}$, i.e., $K \in \{5, 6, 7, 8, 9, 10\}$. More specifically, if $M_r=2$, $K=5$, else if $M_r=3$, $K=6$, and so forth. DOAs, DODs and path-losses are generated in the same way as in Fig. \ref{fig:nmsevsk}. Fig. \ref{fig:nmsevsmr} shows the result, from which we can see that RICE and RICER work well. We find that in most cases, OMP can successfully resolve several paths but not all of them. It frequently mis-estimates one or two paths, which ultimately leads to unsatisfactory overall performance. Note that when $M_r=3$, the maximum number of paths that the proposed methods can deal with is 6 which equals to the number of paths in this simulation. This validates the correctness of the identifiability analysis in Theorem 2. 

\begin{figure}
	\centering
	\subfigure[$K=3$]{\includegraphics[width=0.9\linewidth]{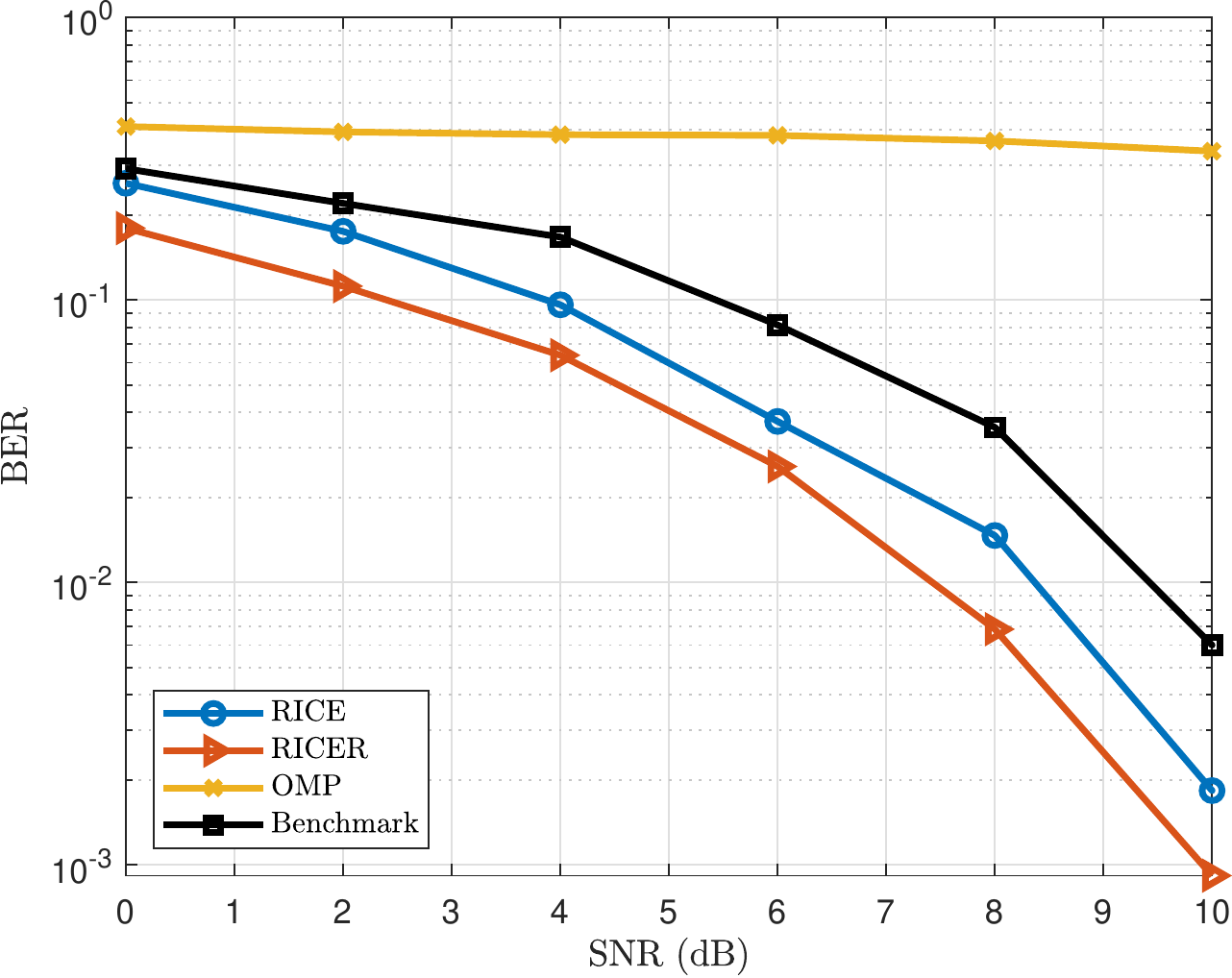}}\label{fig:bervssnrk3mr3}
	\subfigure[$K=4$] {\includegraphics[width=0.9\linewidth]{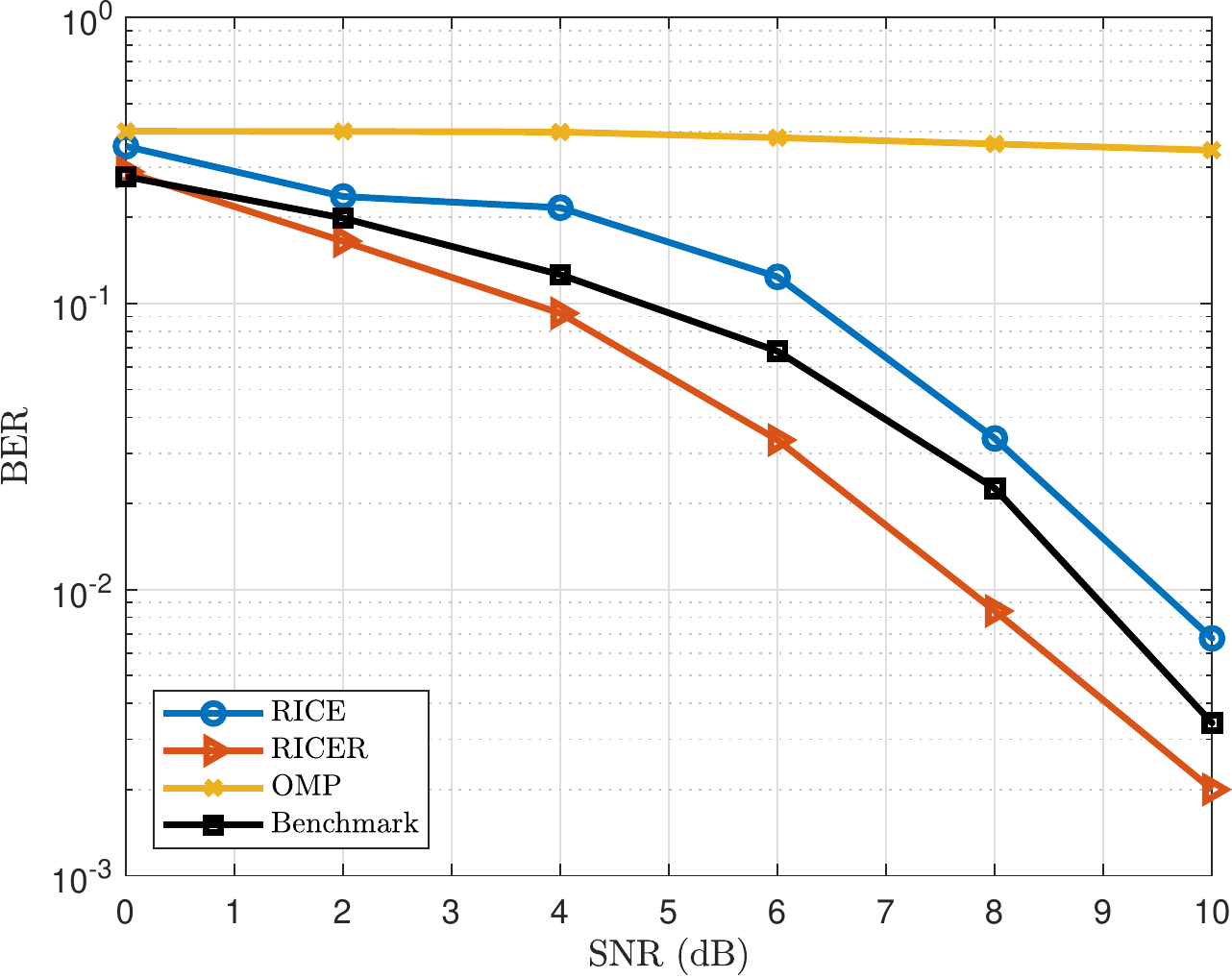}}\label{fig:bervssnrk4mr3}
	\caption{BER versus SNR.} \label{fig:ber}
\end{figure}

In the last example, we examine the channel estimation performance by evaluating the bit error rate (BER) versus SNR. In the simulation, we first estimate the channel and then feed it back to the BS. Then we transmit QPSK symbols precoded using the zero-forcing precoding technique. We pass the coded signal through a white Gaussian channel and decode it at the MS. To simplify the analysis, we do not consider quantization error at the feedback step. The parameters are set as $M_r=3$, $N_x=N_y=2$ and $M_x=M_y=10$. According to Theorem 2, we can uniquely identify up to $4$ paths based on the setting. We consider two cases: $K=3$ and $K=4$. The results are plotted in Fig. \ref{fig:ber}. Note that the benchmark curve is based on the least squares channel estimator with orthogonal pilots. We see that when $K=3$, RICE and RICER outperform the benchmark and OMP. But the RICE algorithm performs worse than the benchmark when $K=4$. We note that the benchmark achieving such performance is at the expense of huge  training and feedback overhead, where the downlink training is based on a $100\times 100$ training signal and uplink feedback is with 600 real-valued numbers. However, the overhead of RICE and RICER is much lighter, where they only spend 16\% of the training overhead of the benchmark and approximately 3.3\% of the feedback overhead.

\section{Conclusion}
In this work, we designed a new non-orthogonal training sequence and proposed a novel tensor factorization framework to tackle the DL channel estimation problem for FDD massive MIMO from `frugal' training. We showed that with the devised training sequence, the channel can be estimated accurately from a very small amount of training. Meanwhile, two computationally efficient algebraic methods were proposed for multipath parameter estimation. 
Compared to the existing approaches, the proposed methods have several advantages in terms of channel identification guarantees, estimation accuracy and computational complexity. 
Extensive simulations showed the effectiveness of the proposed methods. The most important take-away point is that RICE achieves similar or better performance than orthogonal training with a much shorter training sequence and using a computationally very attractive algebraic channel identification algorithm.

\bibliographystyle{IEEEtran}
\bibliography{IEEEabrv,mybib}

\begin{thebibliography}{10}
\providecommand{\url}[1]{#1}
\csname url@samestyle\endcsname
\providecommand{\newblock}{\relax}
\providecommand{\bibinfo}[2]{#2}
\providecommand{\BIBentrySTDinterwordspacing}{\spaceskip=0pt\relax}
\providecommand{\BIBentryALTinterwordstretchfactor}{4}
\providecommand{\BIBentryALTinterwordspacing}{\spaceskip=\fontdimen2\font plus
\BIBentryALTinterwordstretchfactor\fontdimen3\font minus
  \fontdimen4\font\relax}
\providecommand{\BIBforeignlanguage}[2]{{%
\expandafter\ifx\csname l@#1\endcsname\relax
\typeout{** WARNING: IEEEtran.bst: No hyphenation pattern has been}%
\typeout{** loaded for the language `#1'. Using the pattern for}%
\typeout{** the default language instead.}%
\else
\language=\csname l@#1\endcsname
\fi
#2}}
\providecommand{\BIBdecl}{\relax}
\BIBdecl

\bibitem{QiFuSi:SPAWC2019}
C.~Qian, X.~Fu, and N.~D. Sidiropoulos, ``A simple algebraic channel estimation
  method for \uppercase{FDD} massive \uppercase{MIMO} systems,'' in \emph{IEEE
  Int. Workshop on Signal Process. Advances in Wireless Commun. (SPAWC)},
  Cannes, France, July 2019, accepted.

\bibitem{larsson2014massive}
E.~G. Larsson, O.~Edfors, F.~Tufvesson, and T.~L. Marzetta, ``Massive
  \uppercase{MIMO} for next generation wireless systems,'' \emph{{IEEE} Commun.
  Mag.}, vol.~52, no.~2, pp. 186--195, 2014.

\bibitem{heath2016overview}
R.~W. Heath, N.~Gonzalez-Prelcic, S.~Rangan, W.~Roh, and A.~M. Sayeed, ``An
  overview of signal processing techniques for millimeter wave \uppercase{MIMO}
  systems,'' \emph{{IEEE} J. Sel. Topics Signal Process.}, vol.~10, no.~3, pp.
  436--453, 2016.

\bibitem{alkhateeb2014channel}
A.~Alkhateeb, O.~El~Ayach, G.~Leus, and R.~W. Heath, ``Channel estimation and
  hybrid precoding for millimeter wave cellular systems,'' \emph{{IEEE} J. Sel.
  Areas Commun.}, vol.~8, no.~5, pp. 831--846, 2014.

\bibitem{qc2018}
C.~Qian, X.~Fu, N.~D. Sidiropoulos, and Y.~Yang, ``Tensor-based channel
  estimation for dual-polarized massive \uppercase{MIMO} systems,''
  \emph{{IEEE} Trans. Signal Process.}, vol.~66, no.~24, pp. 6390--6403, Dec
  2018.

\bibitem{3gpp_stand}
A.~Kammoun, H.~Khanfir, Z.~Altman, M.~Debbah, and M.~Kamoun, ``Preliminary
  results on 3d channel modeling: From theory to standardization,''
  \emph{{IEEE} J. Sel. Areas Commun.}, vol.~32, no.~6, pp. 1219--1229, 2014.

\bibitem{xie2016overview}
H.~Xie, F.~Gao, and S.~Jin, ``An overview of low-rank channel estimation for
  massive \uppercase{MIMO} systems,'' \emph{{IEEE} Access}, vol.~4, pp.
  7313--7321, 2016.

\bibitem{jarvis}
W.~U. Bajwa, J.~Haupt, A.~M. Sayeed, and R.~Nowak, ``Compressed channel
  sensing: A new approach to estimating sparse multipath channels,''
  \emph{Proc. {IEEE}}, vol.~98, no.~6, pp. 1058--1076, 2010.

\bibitem{van2004optimum}
H.~L. Van~Trees, \emph{Optimum array processing: Part IV of detection,
  estimation, and modulation theory}.\hskip 1em plus 0.5em minus 0.4em\relax
  John Wiley \& Sons, 2004.

\bibitem{stoica1989music}
P.~Stoica and A.~Nehorai, ``Music, maximum likelihood, and
  \uppercase{C}ram\'er-\uppercase{R}ao bound,'' \emph{{IEEE} Trans. Acoust.,
  Speech, Signal Process.}, vol.~37, no.~5, pp. 720--741, 1989.

\bibitem{rao1989performance}
B.~D. Rao and K.~S. Hari, ``Performance analysis of root-\uppercase{MUSIC},''
  \emph{{IEEE} Trans. Acoust., Speech, Signal Process.}, vol.~37, no.~12, pp.
  1939--1949, 1989.

\bibitem{roy1989esprit}
R.~Roy and T.~Kailath, ``Esprit-estimation of signal parameters via rotational
  invariance techniques,'' \emph{{IEEE} Trans. Acoust., Speech, Signal
  Process.}, vol.~37, no.~7, pp. 984--995, 1989.

\bibitem{wax1989unique}
M.~Wax and I.~Ziskind, ``On unique localization of multiple sources by passive
  sensor arrays,'' \emph{{IEEE} Trans. Acoust., Speech, Signal Process.},
  vol.~37, no.~7, pp. 996--1000, 1989.

\bibitem{berger2010application}
C.~R. Berger, Z.~Wang, J.~Huang, and S.~Zhou, ``Application of compressive
  sensing to sparse channel estimation,'' \emph{IEEE Communications Magazine},
  vol.~48, no.~11, pp. 164--174, 2010.

\bibitem{jomp}
X.~Rao and V.~K. Lau, ``Distributed compressive \uppercase{CSIT} estimation and
  feedback for \uppercase{FDD} multi-user massive \uppercase{MIMO} systems,''
  \emph{{IEEE} Trans. Signal Process.}, vol.~62, no.~12, pp. 3261--3271, 2014.

\bibitem{lee2016channel}
J.~Lee, G.-T. Gil, and Y.~H. Lee, ``Channel estimation via orthogonal matching
  pursuit for hybrid \uppercase{MIMO} systems in millimeter wave
  communications,'' \emph{{IEEE} Trans. Wireless Commun.}, vol.~64, no.~6, pp.
  2370--2386, 2016.

\bibitem{he2016pilot}
X.~He, R.~Song, and W.-P. Zhu, ``Pilot allocation for
  distributed-compressed-sensing-based sparse channel estimation in
  \uppercase{MIMO-OFDM} systems,'' \emph{{IEEE} Trans. Veh. Technol.}, vol.~65,
  no.~5, pp. 2990--3004, 2016.

\bibitem{han2019fdd}
Y.~Han, Q.~Liu, C.-K. Wen, S.~Jin, and K.-K. Wong, ``\uppercase{FDD} massive
  \uppercase{MIMO} based on efficient downlink channel reconstruction,''
  \emph{IEEE Transactions on Communications}, 2019.

\bibitem{han2019efficient}
Y.~Han, T.-H. Hsu, C.-K. Wen, K.-K. Wong, and S.~Jin, ``Efficient downlink
  channel reconstruction for \uppercase{FDD} multi-antenna systems,''
  \emph{IEEE Transactions on Wireless Communications}, vol.~18, no.~6, pp.
  3161--3176, 2019.

\bibitem{alevizos2018limited}
P.~N. Alevizos, X.~Fu, N.~D. Sidiropoulos, Y.~Yang, and A.~Bletsas, ``Limited
  feedback channel estimation in massive \uppercase{MIMO} with non-uniform
  directional dictionaries,'' \emph{IEEE Transactions on Signal Processing},
  vol.~66, no.~19, pp. 5127--5141, 2018.

\bibitem{yang2012off}
Z.~Yang, L.~Xie, and C.~Zhang, ``Off-grid direction of arrival estimation using
  sparse bayesian inference,'' \emph{IEEE Transactions on Signal Processing},
  vol.~61, no.~1, pp. 38--43, 2012.

\bibitem{qian2018robust}
C.~Qian, Y.~Shi, L.~Huang, and H.~C. So, ``Robust harmonic retrieval via block
  successive upper-bound minimization,'' \emph{{IEEE} Trans. Signal Process.},
  vol.~66, no.~23, pp. 6310--6324, 2018.

\bibitem{shen2015joint}
W.~Shen, L.~Dai, B.~Shim, S.~Mumtaz, and Z.~Wang, ``Joint \uppercase{CSIT}
  acquisition based on low-rank matrix completion for \uppercase{FDD} massive
  \uppercase{MIMO} systems,'' \emph{{IEEE} Commun. Lett.}, vol.~19, no.~12, pp.
  2178--2181, 2015.

\bibitem{li2018millimeter}
X.~Li, J.~Fang, H.~Li, and P.~Wang, ``Millimeter wave channel estimation via
  exploiting joint sparse and low-rank structures,'' \emph{{IEEE} Trans.
  Wireless Commun.}, vol.~17, no.~2, pp. 1123--1133, 2018.

\bibitem{hugl2002spatial}
K.~Hugl, K.~Kalliola, and J.~Laurila, ``Spatial reciprocity of uplink and
  downlink radio channels in \uppercase{FDD} systems,'' in \emph{Proc. COST},
  vol. 273, no.~2.\hskip 1em plus 0.5em minus 0.4em\relax Citeseer, 2002, p.
  066.

\bibitem{xie2017unified}
H.~Xie, F.~Gao, S.~Zhang, and S.~Jin, ``A unified transmission strategy for
  \uppercase{TDD}/\uppercase{FDD} massive \uppercase{MIMO} systems with spatial
  basis expansion model.'' \emph{{IEEE} Trans. Veh. Technol.}, vol.~66, no.~4,
  pp. 3170--3184, 2016.

\bibitem{liu2002almost}
X.~Liu and N.~D. Sidiropoulos, ``Almost sure identifiability of constant
  modulus multidimensional harmonic retrieval,'' \emph{{IEEE} Trans. Signal
  Process.}, vol.~50, no.~9, pp. 2366--2368, 2002.

\bibitem{liu2}
J.~Liu and X.~Liu, ``An eigenvector-based approach for multidimensional
  frequency estimation with improved identifiability,'' \emph{{IEEE} Trans.
  Signal Process.}, vol.~54, no.~12, pp. 4543--4556, 2006.

\bibitem{nion2010tensor}
D.~Nion and N.~D. Sidiropoulos, ``Tensor algebra and multidimensional harmonic
  retrieval in signal processing for \uppercase{MIMO} radar,'' \emph{{IEEE}
  Trans. Signal Process.}, vol.~58, no.~11, pp. 5693--5705, 2010.

\bibitem{unitaryesprit}
M.~Haardt, F.~Roemer, and G.~Del~Galdo, ``Higher-order \uppercase{SVD}-based
  subspace estimation to improve the parameter estimation accuracy in
  multidimensional harmonic retrieval problems,'' \emph{{IEEE} Trans. Signal
  Process.}, vol.~56, no.~7, pp. 3198--3213, 2008.

\bibitem{smoothESPRIT}
M.~S{\o}rensen and L.~De~Lathauwer, ``Blind signal separation via tensor
  decomposition with \uppercase{V}andermonde factor: Canonical polyadic
  decomposition,'' \emph{{IEEE} Trans. Signal Process.}, vol.~61, no.~22, pp.
  5507--5519, 2013.

\bibitem{sorensen2017multidimensional}
------, ``Multidimensional harmonic retrieval via coupled canonical polyadic
  decomposition—\uppercase{P}art \uppercase{II}: Algorithm and multirate
  sampling,'' \emph{{IEEE} Trans. Signal Process.}, vol.~65, no.~2, pp.
  528--539, 2017.

\bibitem{hua1990matrix}
Y.~Hua and T.~K. Sarkar, ``Matrix pencil method for estimating parameters of
  exponentially damped/undamped sinusoids in noise,'' \emph{{IEEE} Trans.
  Acoust., Speech, Signal Process.}, vol.~38, no.~5, pp. 814--824, 1990.

\bibitem{nehorai1991direction}
A.~Nehorai, D.~Starer, and P.~Stoica, ``Direction-of-arrival estimation in
  applications with multipath and few snapshots,'' \emph{Circuits, Systems and
  Signal Processing}, vol.~10, no.~3, pp. 327--342, 1991.

\bibitem{Sid2017}
N.~D. Sidiropoulos, L.~De~Lathauwer, X.~Fu, K.~Huang, E.~E. Papalexakis, and
  C.~Faloutsos, ``Tensor decomposition for signal processing and machine
  learning,'' \emph{{IEEE} Trans. Signal Process.}, vol.~65, no.~13, pp.
  3551--3582, 2017.

\bibitem{khabbazibasmenj2014efficient}
A.~Khabbazibasmenj, A.~Hassanien, S.~A. Vorobyov, and M.~W. Morency,
  ``Efficient transmit beamspace design for search-free based \uppercase{DOA}
  estimation in \uppercase{MIMO} radar.'' \emph{{IEEE} Trans. Signal Process.},
  vol.~62, no.~6, pp. 1490--1500, 2014.

\bibitem{fu2015factor}
X.~Fu, N.~D. Sidiropoulos, J.~H. Tranter, and W.-K. Ma, ``A factor analysis
  framework for power spectra separation and multiple emitter localization,''
  \emph{IEEE Transactions on Signal Processing}, vol.~63, no.~24, pp.
  6581--6594, 2015.

\bibitem{schmidt1986multiple}
R.~Schmidt, ``Multiple emitter location and signal parameter estimation,''
  \emph{{IEEE} Trans. Antennas Propag.}, vol.~34, no.~3, pp. 276--280, 1986.

\bibitem{shaghaghi2015subspace}
M.~Shaghaghi and S.~A. Vorobyov, ``Subspace leakage analysis and improved
  \uppercase{DOA} estimation with small sample size.'' \emph{{IEEE} Trans.
  Signal Process.}, vol.~63, no.~12, pp. 3251--3265, 2015.

\end{thebibliography}

\end{document}